\documentclass[a4paper,12pt]{article}
\pdfoutput=1
\usepackage{epsfig}
\usepackage{amssymb,subfigure}
\usepackage{amsfonts}
\usepackage{amsmath}
\usepackage{euscript}
\usepackage{verbatim}
\usepackage{latexsym}
\usepackage{graphicx,color}
\usepackage{caption}
\usepackage{float}
\usepackage{slashed}
\usepackage{graphicx}
\graphicspath{ {images/} }
\usepackage{ytableau}

\usepackage{multirow}

\usepackage{mathtools}

\usepackage[colorinlistoftodos]{todonotes}
\usepackage[colorlinks=true, allcolors=blue]{hyperref}

\usepackage{wrapfig}
	\usepackage[T1]{fontenc}
	\usepackage{tikz}
	\usetikzlibrary{decorations.pathmorphing}
\usepackage{tikz}
\usetikzlibrary{shapes.geometric, arrows,patterns,snakes}
\tikzstyle{ellip} = [ellipse, minimum width=3cm, minimum height=1cm,text centered, draw=black]

\newskip\humongous \humongous=0pt plus 1000pt minus 1000pt

\newif\ifdtup

\allowdisplaybreaks[1]%To allow page-breaks while using align%

\catcode`\@=11

\@addtoreset{equation}{section}

\def\@normalsize{\@setsize\normalsize{15pt}\xiipt\@xiipt
\abovedisplayskip 14pt plus3pt minus3pt%
\belowdisplayskip \abovedisplayskip
\abovedisplayshortskip \z@ plus3pt%
\belowdisplayshortskip 7pt plus3.5pt minus0pt}

\def\small{\@setsize\small{13.6pt}\xipt\@xipt
\abovedisplayskip 13pt plus3pt minus3pt%
\belowdisplayskip \abovedisplayskip
\abovedisplayshortskip \z@ plus3pt%
\belowdisplayshortskip 7pt plus3.5pt minus0pt
\def\@listi{\parsep 4.5pt plus 2pt minus 1pt
     \itemsep \parsep
     \topsep 9pt plus 3pt minus 3pt}}

\relax

\catcode`@=12

\catcode`\@=11

\def\section{\@startsection{section}{1}{\z@}{3.5ex plus 1ex minus
   .2ex}{2.3ex plus .2ex}{\large\bf}}

\def\SymBoxes#1#2#3#4{\newdimen\un@t \un@t#3%
\raisebox{#1}{\rule{#2\un@t}{#4}\hskip-#2\un@t% lower horizontal
\@tempdimb\un@t \advance\@tempdimb by-#4\@tempcntb#2\relax%
\@whilenum{\@tempcntb>0}\do{%                         % #2 vertical lines
\rule{#4}{\un@t}\hskip\@tempdimb \advance\@tempcntb by\m@ne}%
\hskip-#2\un@t \rule[\un@t]{#2\un@t}{#4}%
\rule[\un@t]{#4}{#4}\hskip-#4%             % upper horizontal line
\rule{#4}{\un@t}}\hskip-#4}                % rightest vertical line
\begin{document}
%\begin{letter}{~}

%%%%%%Define some new commands and  macros
\newcommand{\beq}{\begin{equation}}
\newcommand{\eeq}{\end{equation}}
\newcommand{\bea}{\begin{eqnarray}}
\newcommand{\eea}{\end{eqnarray}}
\newcommand{\beas}{\begin{eqnarray*}}
\newcommand{\eeas}{\end{eqnarray*}}
\newcommand{\defi}{\stackrel{\rm def}{=}}
\newcommand{\non}{\nonumber}
\newcommand{\bquo}{\begin{quote}}
\newcommand{\enqu}{\end{quote}}
%%%%%%%%%%%%%%%%
\renewcommand{\(}{\begin{equation}}
\renewcommand{\)}{\end{equation}}
%%%%%%%%%%%%%%%%%%%%%%%%%%%%%%%%%% definitions
\def \eqn#1#2{\begin{equation}#2\label{#1}\end{equation}}
\def\IZ{{\mathbb Z}}
\def\IR{{\mathbb R}}
\def\IC{{\mathbb C}}
\def\IQ{{\mathbb Q}}
\def\de{\partial}
\def\Tr{ \hbox{\rm Tr}}
\def\H{ \hbox{\rm H}}
\def\HE{ \hbox{$\rm H^{even}$}}
\def\HO{ \hbox{$\rm H^{odd}$}}
\def\K{ \hbox{\rm K}}
\def\Im{ \hbox{\rm Im}}
\def\Ker{ \hbox{\rm Ker}}
\def\const{\hbox {\rm const.}}
\def\o{\over}
\def\im{\hbox{\rm Im}}
\def\re{\hbox{\rm Re}}
\def\bra{\langle}\def\ket{\rangle}
\def\Arg{\hbox {\rm Arg}}
\def\Re{\hbox {\rm Re}}
\def\Im{\hbox {\rm Im}}
\def\exo{\hbox {\rm exp}}
\def\diag{\hbox{\rm diag}}
\def\longvert{{\rule[-2mm]{0.1mm}{7mm}}\,}
\def\a{\alpha}
\def\dag{{}^{\dagger}}
\def\tq{{\widetilde q}}
\def\p{{}^{\prime}}
\def\W{W}
\def\N{{\cal N}}
\def\hsp{,\hspace{.7cm}}

\def\br{\nonumber\\}
\def\IZ{{\mathbb Z}}
\def\IR{{\mathbb R}}
\def\IC{{\mathbb C}}
\def\IQ{{\mathbb Q}}
\def\IP{{\mathbb P}}
\def \eqn#1#2{\begin{equation}#2\label{#1}\end{equation}}

\newcommand{\sgm}[1]{\sigma_{#1}}
\newcommand{\idd}{\mathbf{1}}

\newcommand{\C}{\ensuremath{\mathbb C}}
\newcommand{\Z}{\ensuremath{\mathbb Z}}
\newcommand{\R}{\ensuremath{\mathbb R}}
\newcommand{\rp}{\ensuremath{\mathbb {RP}}}
\newcommand{\cp}{\ensuremath{\mathbb {CP}}}
\newcommand{\vac}{\ensuremath{|0\rangle}}
\newcommand{\vact}{\ensuremath{|00\rangle}                    }
\newcommand{\oc}{\ensuremath{\overline{c}}}
\begin{titlepage}
\begin{flushright}
CHEP XXXXX
%ULB-TH/09-10\\
%hep-th/yymmnnn\\
\end{flushright}
\bigskip
\def\thefootnote{\fnsymbol{footnote}}

\begin{center}
{\Large
{\bf Bounds on Slow Roll at the Boundary of the Landscape  %: \\ The Case of Tree Level Type II String Theory\\
%\vspace{0.2in}
%in Six Dimensions 
}
}
\end{center}

%\bigskip
\begin{center}
{Sumit K. GARG$^{a}$,\footnote{\texttt{sumit.k@cmr.edu.in}} \ \ Chethan KRISHNAN$^b$,\footnote{\texttt{chethan.krishnan@gmail.com}} \ \ M. Zaid ZAZ$^c$ \footnote{\texttt{mohammadzaz@gmail.com}}%\vspace{0.15in} 
\\ }
%\vspace{0.1in}

\end{center}

\renewcommand{\thefootnote}{\arabic{footnote}}

\begin{center}
%\vspace{0.2cm}
$^a$ {Department of Physics, \\
CMR University, Bengaluru 562149, India }

$^b$ {Center for High Energy Physics,\\
Indian Institute of Science, Bangalore 560012, India}

$^c$ {International Center for Theoretical Sciences,\\
Tata Institute of Fundamental Research, Bangalore 560089, India}

\end{center}

\noindent
\begin{center} {\bf Abstract} \end{center}
We present strong evidence that the tree level slow roll bounds of arXiv:1807.05193 and arXiv:1810.05506 are valid, even when the tachyon has overlap with the volume of the cycle wrapped by the orientifold. This extends our previous results in the volume-dilaton subspace to a semi-universal modulus. Emboldened by this and other observations, we investigate what it means to have a bound on (generalized) slow roll in a multi-field landscape. We argue that for {\em any} point $\phi_0$ in an $N$-dimensional field space with $V(\phi_0) > 0$, there exists a path of monotonically decreasing potential energy to a point $\phi_1$  within a path length $\lesssim {\cal O}(1)$, such that  $\sqrt{N}\ln \frac{V(\phi_1)}{V(\phi_0)} \lesssim - {\cal O} (1)$. The previous de Sitter swampland bounds are specific ways to realize this stringent non-local constraint on field space, but we show that it also incorporates (for example) the scenario where both slow roll parameters are intermediate-valued and the Universe undergoes a small number of e-folds, as in the Type IIA set up of arXiv:1310.8300. Our observations are in the context of tree level constructions, so we take the conservative viewpoint that it is a characterization of the classical "boundary" of the string landscape. To emphasize this, we argue that these bounds can be viewed as a type of Dine-Seiberg statement.

\vspace{1.6 cm}
\vfill

\end{titlepage}

\setcounter{page}{2}
%\tableofcontents

\setcounter{footnote}{0}

%%%%%%%%%%%%%%%%%%%%%%%%%%%%%%%%%%%%%%%%%%%%%%%%%%%%%%%%%%%%%%%%%%%%%%%%%%%%%%%%%%%%%%%%%%%%%%
%%%%%%%%%%%%%%%%%%%%%%%%%%%%%%%%%%%%%%%%%%%%%%%%%%%%%%%%%%%%%%%%%%%%%%%%%%%%%%%%%%%%%%%%%%%%%%

%\section{Introduction}

\section{De Sitter Denialism at the Boundary of the Landscape}

Fully explicit and controllable (meta-)stable de Sitter vacua in string theory have been a challenge to construct \cite{Carlos, Ulf, Sethi, VafaReview, many}, with various moving parts that have raised controversy. Based on this, it has recently been conjectured that it is impossible to realize slow roll\footnote{Unless otherwise explicitly stated, by ``slow roll" we will always mean what is sometimes referred to as ``potential slow roll'' defined via the $\epsilon_V$ and $\eta_V$ parameters of the potential. We will say more about the precise notion of slow roll we are trying to capture, in later sections.} in string theory \cite{Vafa1, GK}. In this paper, we will strengthen the evidence for some of these claims, and seek a broader context in which to  view them. We start by emphasizing that all the evidence from string theory for these and previous {\em de Sitter Swampland} bounds are found in the context of tree level string/M-theory with singular sources. Therefore in this paper, we will conservatively view our observations as statements about the classical ``boundary" of the string landscape. The reader can decide whether he/she would like to view these as evidence for the absence of de Sitter in all of string theory or merely as a characterization of the boundary of the landscape. 

%However, there exist reasonable (in our view anyway) arguments based on effective field theory \cite{Joe, Shamit-Sandip} that suggest that the de Sitter constructions in the full {\em quantum} string theory are indeed viable. In this paper, we will try to reconcile these two facts by presenting a perspective on how to understand the de Sitter swampland conjectures.

%It is {\em not} obvious to us that string theory does {\em not} have de Sitter vacuua. 

Before we proceed, let us state some of our prejudices. We do not find it immediately problematic that quantum corrections seem always necessary for constructing de Sitter vacua in string theory. As has often been emphasized, if we turn off quantum corrections, atoms also do not exist\footnote{A fact that is often {\em not} emphasized however is that because of dualities in string theory, the statement about de Sitter is in fact stronger. The classical limit in one duality frame contains the quantum effects in another. However we do not believe this significantly alters the punchline, because if one requires finite quantum corrections in one duality frame to get de Sitter, that fact remains invariant under dualities.}. One way to think about this is to notice an old result by Farhi and Guth \cite{Farhi-Guth}, where they showed that de Sitter cannot emerge classically (ie., without a singularity) in an asymptotically flat space due to Penrose's singularity theorem\footnote{Perhaps unsurprising because unlike flat space and Anti-de Sitter, de Sitter does not have a spatial asymptotic region.}. This statement is independent of string theory, so it is unsurprising\footnote{Let us emphasize however that we are not aware of a general {\em a-priori} argument in the literature, that tree level constructions cannot lead to parametrically controlled de Sitter vacua. The fact that a huge amount of effort has been invested into this, and yet so far there is not a single de Sitter vacuum that is parametrically controlled, nor a general {\em a-priori} understanding of what might be the problem, is striking. The suggestion that this is not a problem because these are tree level constructions, is somewhat cavalier: why then has there been a large number of papers trying to look for precisely such vacua?} that classical string constructions do not find de Sitter. The non-trivial question in our view, is to know whether a consistent de Sitter space can exist at all in a quantum theory. If we have independent reasons (other than string constructions or lack thereof) to believe that de Sitter can exist or not in a quantum theory, then that would clarify things quite a bit. What is worrisome in our view is not that de Sitter seems to require quantum effects in string theory, but that whenever things are explicitly calculable, a sufficiently stable (quasi-)de Sitter space seems to slip through our hands in string theory. To be fair however, typically such fully calculable constructions use only tree level ingredients, even though there does exist one No-Go theorem that we are aware of which forbids de Sitter even with a certain limited class of quantum corrections \cite{Quigley}. 

Given these facts, what we hope can emerge out of the current discussion is a clear {\em a-priori} statement about what kind of effects are necessary to produce de Sitter, if at all it is possible to do that in string theory. A conservative starting point to think about the present situation is that de Sitter can only be found deep inside the thick of the moduli space \cite{Dine-Seiberg, Polchinski}, and not at the boundary, and we would like to know what precisely characterizes this boundary\footnote{This is to be compared to how classical physics can be viewed as the $\hbar \rightarrow 0$ boundary of the semi-infinite line $\hbar \in [0, \infty)$. One can make $\hbar$ dimensionless if one wants by considering a one-parameter family of solutions.}. % with a dimensionful paramter, and then the analogy with the string coupling (say) becomes tighter.}. 
To do this, it could be useful to sharpen the conjectures of \cite{Vafa1, GK} as much as we can, and this is one of our motivations in the present paper. We expect that our statements should hold for either all UV complete theories (in which case a de Sitter vacuum is ruled out in string theory) or at the very least for those UV complete theories that are at the classical limit of the moduli space (in which case de Sitter could exist in the interior, and our approach is useful for charting out the boundary of the landscape). 

In particular, in this paper we will generalize the conjectures of \cite{Vafa1, GK}, and formulate a statement that we suspect is consistent with all known tree level constructions with sources in string/M-theory. We will start by reviewing the bound in \cite{GK} and discussing the directions in which it will be developed in this paper.

{\bf Comment Added:} While this paper was in its final stages, \cite{Return-of-the-Vafa} appeared. The ``Refined de Sitter Conjecture'' of that paper is identical to the statement in the abstract of \cite{GK} ``that $\epsilon_V$ need not necessarily be ${\cal O}(1)$ if $\eta_V \lesssim -{\cal O}(1)$ holds''. This slow roll bound is the starting point of the present paper, and among other things, we will further generalize it. We will discuss situations where the above bound is in tension, but the generalized bound holds without qualifications.

\subsection{Slow Roll Bounds}

The statement of \cite{Vafa1} (see also \cite{Vafa2}) is a statement about the (magnitude of the) slope of the potential. For positive values of the potential, it implies that at any point in the landscape, one should expect the magnitude of the slope of the potential to be about equal or bigger than the value of the potential itself, measured in Planck units. The argument generalizes previous No-Go statements in the literature \cite{HKTT} into a general principle.

As it stands however, this statement has counter-examples \cite{Andriot, GK} in tree level type II flux compactifications, where solutions with zero slope at positive values of the potential have already been constructed (see, eg., \cite{Koerber}). 
%Furthrmore, a bottom-up approach was taken recently in \cite{Denef}, who noted that in a Universe with a quintessence field and a Higgs field, the bound of \cite{Vafa1} can be violated by fifty five orders of magnitude. A similar comment can be made about the situation pointed out in \cite{Denef} which also relied on the peak of the Higgs potential.  
However, all such type II de Sitter solutions have at least one tachyonic direction \cite{Koerber, Ulf}, and in fact general arguments have been raised in \cite{Shiu, VanRiet} to show that in wide classes of such potential de Sitter solutions, tachyons are inevitable. In other words, one is left with the impression that these solutions violate the letter, but not the spirit, of the bound proposed by \cite{Vafa1}. Is there a way to make this intuition precise? In other words, is there a plausible variation of the conjecture in \cite{Vafa1} which could successfully incorporate these ``counter-examples"? 

The main idea presented in \cite{GK} was that both these questions are naturally answered, if one views the bound as an ${\cal O}(1)$ bound on slow roll. In particular, it was suggested that in situations where the first slow roll parameter $\epsilon_V$ is zero (or very close to zero), the second slow roll parameter $\eta_V$ must be $\lesssim -{\cal O}(1)$. Indeed, by a systematic analysis that took advantage of the scaling behavior of the potential terms, it was shown in \cite{GK} that the classes of tachyons found in \cite{Shiu, VanRiet} all satisfy such an ${\cal O}(1)$ bound.   %This was presented as evidence for such a slow roll bound, in \cite{GK}. 
It was also pointed out there that the masses of all the listed tachyons in explicit string theory constructions in the literature, also satisfy such a bound\footnote{There are a few papers that appeared after the appearance of \cite{GK} that note that the original \cite{Vafa1} proposal runs into trouble in some situations that involve a saddle or peak of the effective potential: these include \cite{Denef,Conlon,Yamazaki} and possibly others. These objections automatically go away when the refined bound is used. It will be interesting to see if these phenomenology-inspired arguments can be used as evidence that these bounds should hold beyond tree level string theory (as we tend to suggest in this paper).}. 

One point that was left unaddressed in \cite{GK} was that the discussion there was largely limited to the case where the tachyon was in a certain universal subspace of the moduli space. This was the subspace spanned by the dilaton and the volume modulus of the compact manifold. If the tachyon falls in that subspace, the bounds are powerful as was shown in \cite{GK}, but if the tachyon is not in that subspace, the entire discussion is moot. This becomes especially important because in large classes of de Sitter constructions, eg.  \cite{obstinate}, it is known that the tachyon can actually be outside this subspace. What can be said about the bound in such a situation? In this paper, we will address this issue for a large class of solutions listed in \cite{obstinate} where the tachyon has overlap with the orientifold cycle volume, which is a semi-universal modulus. We will present strong evidence by combining semi-analytic methods and  computerized scans that indeed, such tachyons also satisfy the bound. 
 
While the basic idea of a slow roll bound was qualitatively stated in \cite{GK}, a precise statement was only given in two situations: when $\epsilon_V$ is large (and the bound is satisfied as in \cite{Vafa1}), or when $\epsilon_V$ is zero or exceedingly small and $|\eta_V|$ is large (and the bound is satisfied as in \cite{GK}). While this was sufficient for investigating the situations discussed in \cite{GK}\footnote{In particular, to bound the tachyon masses, one could work specifically at (or very near) the critical point where $\epsilon_V=0 $.}, it is clearly beneficial to have a statement about the bound when the value of $\epsilon_V$ and $\eta_V$ are in some appropriate sense, intermediate-valued. More to the point, we would like to have a better understanding of the question: {\em what feature of the underlying potential is the slow roll bound capturing?} In  other words, is there a more elegant way to phrase the bound as a statement about the potential and not get caught up in special cases having to do with the sizes of $\epsilon_V$ and/or $\eta_V$?  In yet other words, we wish to understand what meaning could be attributed to a bound on slow roll in a general potential landscape. We will argue that a natural interpretation for such a bound is as a non-local statement in field space regarding the values of the potential. More concretely, we will argue that a landscape with a slow roll bound is one in which for any point $\phi_0$ in field space with $V(\phi_0) > 0$, there exists a point $\phi_1$ within a field distance ${\cal O}(1)$, such that  
\bea
\sqrt{N} \ln \frac{V(\phi_1)}{V(\phi_0)} \lesssim - {\cal O} (1).
\eea 
In fact it turns out that a somewhat stronger version of this statement is what emerges naturally from considerations of the slow roll bounds of \cite{GK}: there exists a path connecting $\phi_0$ and such a $\phi_1$ along which the potential is monotonically decreasing. Let us emphasize right away that our demand is {\em not} merely that the absolute value of the left hand side be bounded\footnote{Even though in the concrete examples in Type II that we discussed in a related context in \cite{GK}, the numerical value of the bound was the non-trivial part, and therefore was emphasized.}. This is because such a bound would allow de Sitter vacua, which we will view as the slowest form of slow roll there is! We are seeking a natural way to bound slow roll {\em away} from ``slowness'', so allowing de Sitter would hardly be a promising place to start. 

Note that if there is an extra demand that the potential is positive, our bound can give a simple understanding of the smallness of the cosmological constant. There is an ever-present way down the landscape from any point in field space, and if the potential is bounded from below by zero, eventually it will reach arbitrarily small values of vacuum energy. But of course, it is not very clear how to make sense of trans-Planckian field ranges, so we will not emphasize this point.

Let us also emphasize that the question of defining a slow roll bound is not merely of aesthetic value, it also has practical implications. Indeed, there do exist Type IIA string constructions in the literature \cite{Dibitetto} where $\epsilon_V \sim 0.1$ and $|\eta_V| \sim  0.1$ simultaneously, and such solutions exhibit a few e-folds. It is not very clear from the discussions so far in the literature whether this is in conflict with any of the bounds. We will show by an explicit consideration of the 4D effective potential of that set up, that the potential is indeed comfortably consistent with the version of the bound we present in this paper. The key point, we will see, is that unlike in other scenarios, the direction of the ${\cal O}(1)$ fall in the potential does not overlap with any of the elementary field directions\footnote{In more generic situations, this happens quite generically for many field directions.} and is in fact fairly non-trivial to find. Another interesting feature of this class of solutions is that the field space is 14-dimensional, and we can clearly see the relevance of the $\sqrt{N}$ in making sure that the bound is satisfied.  

Our perspective in this paper is largely quite conservative. The bounds we find are for tree level string constructions, so we will claim the validity of the bound only at the classical ``boundary" of the string landscape\footnote{But we note that such tree level evidence has been used to argue in \cite{Vafa1, Vafa2} that the bounds hold more generally in the string landscape.}. We will in fact invoke a Dine-Seiberg like argument to suggest that the swampland bounds should naturally arise in the perturbative limit of string theory. A closely related question is that of parametric control in de Sitter constructions. The idea here is that even if any given string construction of de Sitter must be at a finite value of the string coupling due to Dine-Seiberg, there can exist parametric families of solutions where by tuning the parameter, one can go to arbitrarily weak coupling. 
No one has been able to construct such parametrically controlled de Sitter vacua in string theory, even in tree level constructions\footnote{We thank Shamit Kachru for a discussion on this.}$^{,}$\footnote{In this context, it is worth noting the very recent effort to construct non-tachyonic tree level constructions using anti-D branes as an ingredient \cite{Kallosh}. The solutions found so far in this class do not live in parametric families that continue to weak coupling. Even if one were to hope that this situation will change in the future, there is another problem. When the system becomes classical, the possibility arises that the brane-flux annihilation \cite{KPV, CK-Jarah, Van} will become un-suppressed, and that it will become unstable. We thank Thomas Van Riet and Timm Wrase for discussions on related points.}. It will be useful to have an understanding of why such families do not seem to exist.

\section{Bounding Orientifold Tachyons}

In our previous paper \cite{GK} most of the discussion was in the context of the universal moduli: the string coupling and the volume of the compact space. This enabled us to come up with essentially analytic arguments for bounding the $\eta_V$-parameter of the tachyon when it was in this subspace. This is elegant, but it suffers from the disadvantage that often the tachyon has legs outside the volume-dilaton subspace. 

Since de Sitter vacua necessarily require the presence of negative tension sources, a standard ingredient in de Sitter tree level constructions is an orientifold plane that wraps a cycle in the compact manifold. The volume of this cycle is another modulus from the 4D perspective, and often the tachyon has support in that direction. In \cite{obstinate} this modulus was called a semi-universal modulus. Unlike the universal moduli, it is not quite possible to treat the orientifold cycle in absolute generality, and the form of the potential depends on various details of the compactification. 
So what we will do instead here is to consider a specific class of tachyonic de Sitter solutions found in  \cite{obstinate} for compactification on an $SU(3)$ structure manifold, and investigate whether they satisfy the bound. We present a scatter plot of the $\eta_V$ parameters of all the candidate solutions in the figure: the very striking ``wall" in the figure which they never cross, is a strong suggestion that the bound holds. The (numerical) strategy that we adopt to make our case is somewhat simplistic at some points, so we will emphasize them as they arise. Despite these assumptions we make, we believe that the fact that we find a wall is suggestive. But we should emphasize that our result should really be viewed as a type of genericity argument that the bound should hold for these tachyons, and not as a rigorous proof.

The set up we will look at in detail is that of massive Type IIA on $SU(2)\times SU(2)$  containing four intersecting space-filling O6 planes \cite{obstinate}. The effective potential of the 4D theory can be expressed in terms of the Kahler potential
\bea
K=-\log (z_1 + \bar z_1)-3 \log (z_2 +\bar z_2)-3 \log (t+\bar t) + 5 \log 2
\eea
and superpotential 
\bea
W= i t^3+3 t( t+z_1+z_2) -i \lambda (z_1-3 z_2), 
\eea
where $\lambda$ is a (mostly) free real parameter from the 4D perspective. In writing the superpotential we have fixed what we believe is a small typo in \cite{obstinate}. We are also taking the following definitions for the real fields in terms of the holomorphic variables as
\bea
t= \rho+ib, \ \ z_1= \tau \sigma^{-9/2}+ ic_1, \ \ z_2 = \tau \sigma^{3/2}+ic_2
\eea
which also fixes a sign in \cite{obstinate}. Without these corrections, we are unable to reproduce the potentials and various other intermediate expressions that they present. Note further that the corrected superpotential still has the form of  an S-T-U supergravity with linear terms for $S$ and $T$ fields (even though they are named differently here), which is expected in this class of flux compactifications. 

The effective potential one gets from the Kahler potential and superpotential contains $c_1$ and $c_2$ but they can be explicitly and algebraically solved away. This is a major simplification. Plugging them back in to the potential, we get the four dimensional effective potential that we will use in the form presented in \cite{obstinate}: 
\begin{eqnarray}
V_{eff} &=& V_0 + V_2 + V_{R} + V_{O6} + V_H\nonumber\\
&=&  \frac{\rho^3}{\tau^4} + \left \{ \frac{3 b^2 \rho}{\tau^4} +\frac{3\rho}{\tau^4}-\frac{6 b\rho}{\tau^4} \right \}+ \left \{ 
\frac{3}{\rho\tau^2\sigma^9} - \frac{12}{\rho\tau^2\sigma^3}-\frac{3\sigma^3}{\rho\tau^2}\right \}+ \left\{ \frac{2(\lambda-3)}{\tau^3 \sigma^{9/2}}
-\frac{6(\lambda+1)\sigma^{3/2}}{\tau^3} \right \}\nonumber\\
&+& \left \{ \frac{\lambda^2}{\rho^3 \tau^2 \sigma^9} + \frac{9 b^2}{\rho^3 \tau^2 \sigma^9} -\frac{6\lambda b}{\rho^3 \tau^2 \sigma^9} 
+\frac{3\lambda^2 \sigma^3}{\rho^3 \tau^2} + \frac{3 b^2 \sigma^3 }{\rho^3 \tau^2} +\frac{6\lambda b \sigma^3}{\rho^3 \tau^2}  \right \}\nonumber\\
&=& V_0 + V_2^{(1)} +V_{2}^{(2)}-V_{2}^{(3)} + V_R^{(1)}-V_R^{(2)}-V_R^{(3)}+(\lambda-3) V_{O6}^{(1)}-(\lambda+1)V_{O6}^{(2)}+\nonumber\\
 &+& \lambda^2 V_{H}^{(1)}+  V_{H}^{(2)}- \lambda V_{H}^{(3)}
+ \lambda^2 V_{H}^{(4)}+ V_{H}^{(5)}+ \lambda V_{H}^{(6)}
\end{eqnarray}
Our first simplifying choice is that in the following we will work with this potential, and not the one containing the RR-axions $c_1$ and $c_2$. As already emphasized, our approach should be viewed as trying to find evidence for the bound, and not as proof\footnote{Note that critical points are unaffected by this switch between potentials. This is what makes our approach plausible. The second derivatives however {\em are} affected, but since we are trying to bound the most negative eigendirection of the second derivative, if we do find a bound (and we will!) in the reduced potential, we find it reasonable to read that as evidence for such a bound in the un-reduced potential as well. A related issue is that once we go to the reduced potential, we lose our control on the precise form of the Kahler kinetic terms. But these again do not affect the location of the critical points, and as far as the second derivatives are considered, these factors are roughly $O(1)$ numbers in the ranges of field values that we will work with, and unless there is a conspiracy, we expect them to not change the qualitative fact that a bound exists. This is indeed what we find.}. 

%In principle, we should work with the former, but our numerical methods are not very sophisticated, and therefore we find it hard to  make useful progress in scanning for critical points when these  extra fields are also present (for multiple reasons that we will not get into here).

We define as in \cite{GK} quantities  \bea \hat{\rho} \propto \ln\rho,\hspace{1cm} ~ \hat{\tau} \propto \ln\tau,\hspace{1cm} ~ \hat{b} \propto \ln b,\hspace{1cm} ~ \hat{\sigma} \propto \ln\sigma  \label{scalign}\eea.
The de Sitter critical points are determined by the conditions
\begin{eqnarray}
-\frac{\partial V_{eff}}{\partial \hat{\rho}}&=& 3 V_H + V_R -3 V_0-V_2=0 \\
-\frac{\partial V_{eff}}{\partial \hat{\tau}}&=& 2 V_H +2 V_R + 3 V_{O6}+ 4 V_0 + 4V_2=0\\ 
-\frac{\partial V_{eff}}{\partial \hat{b}}&=&  V_2^{(3)}-2 V_2^{(1)}-2 V_H^{(2)}+\lambda V_H^{(3)}-2 V_H^{(5)}-\lambda V_H^{(6)}=0 \\
-\frac{\partial V_{eff}}{\partial \hat{\sigma}}&=& 9 V_R^{(1)}-3 V_R^{(2)}+ 3 V_R^{(3)}
+\frac{9}{2}(\lambda-3)V_{O6}^{(1)}+\frac{3}{2}(\lambda+1)V_{O6}^{(2)}+ 9\lambda^2 V_H^{(1)}+ 9 V_H^{(2)}- 9\lambda V_H^{(3)}\nonumber\\
&-& 3\lambda^2 V_H^{(4)}- 3 V_H^{(5)}- 3\lambda V_H^{(6)}=0
\end{eqnarray}
It is difficult to solve these equations directly on a computer, so instead we will use a scanning strategy. So first we solve the above equations as follows:
\begin{itemize}
\item $V_0=-V_R^{(1)}-\frac{1}{6}V_R^{(2)}-\frac{4}{3}V_R^{(3)}+\frac{7}{3}V_H^{(5)}
+\frac{7}{3}\lambda V_H^{(6)}-\frac{(\lambda-3)}{2}V_{O6}^{(1)}-\frac{2}{3}(\lambda+1)V_{O6}^{(2)}
+\frac{7}{3}\lambda^2 V_H^{(4)}$

\item $V_2^{(1)}=V_R^{(1)}-\frac{1}{3}V_R^{(2)}+\frac{1}{3}V_R^{(3)}+\frac{1}{2}V_2^{(3)}
-\frac{4}{3}V_H^{(5)}-\frac{\lambda}{2}V_H^{(3)}-\frac{5}{6}\lambda V_H^{(6)}+\frac{(\lambda-3)}{2}V_{O6}^{(1)}+\frac{1}{6}(\lambda+1)V_{O6}^{(2)}
+\lambda^2 V_H^{(1)}-\frac{\lambda^2}{3}V_H^{(4)}$

\item $V_2^{(2)}=\frac{5}{6}V_R^{(2)}+\frac{5}{3}V_R^{(3)}+\frac{1}{2}V_2^{(3)}
-\frac{5}{3}V_H^{(5)}+\frac{\lambda}{2}V_H^{(3)}-\frac{13}{6}\lambda V_H^{(6)}-\frac{(\lambda-3)}{2}V_{O6}^{(1)}+\frac{4}{3}(\lambda+1)V_{O6}^{(2)}
-\lambda^2 V_H^{(1)}-\frac{8}{3}\lambda^2 V_H^{(4)}$

\item $V_H^{(2)}=-V_R^{(1)}+\frac{1}{3}V_R^{(2)}-\frac{1}{3}V_R^{(3)}
+\frac{1}{3}V_H^{(5)}+\lambda V_H^{(3)}+\frac{1}{3}\lambda V_H^{(6)}-\frac{(\lambda-3)}{2}V_{O6}^{(1)}-\frac{1}{6}(\lambda+1)V_{O6}^{(2)}
-\lambda^2 V_H^{(1)}+\frac{1}{3}\lambda^2 V_H^{(4)}$
\end{itemize}
\begin{figure}
\centering
\begin{minipage}{\textwidth}
  \centering
  \includegraphics[width=0.9\linewidth]{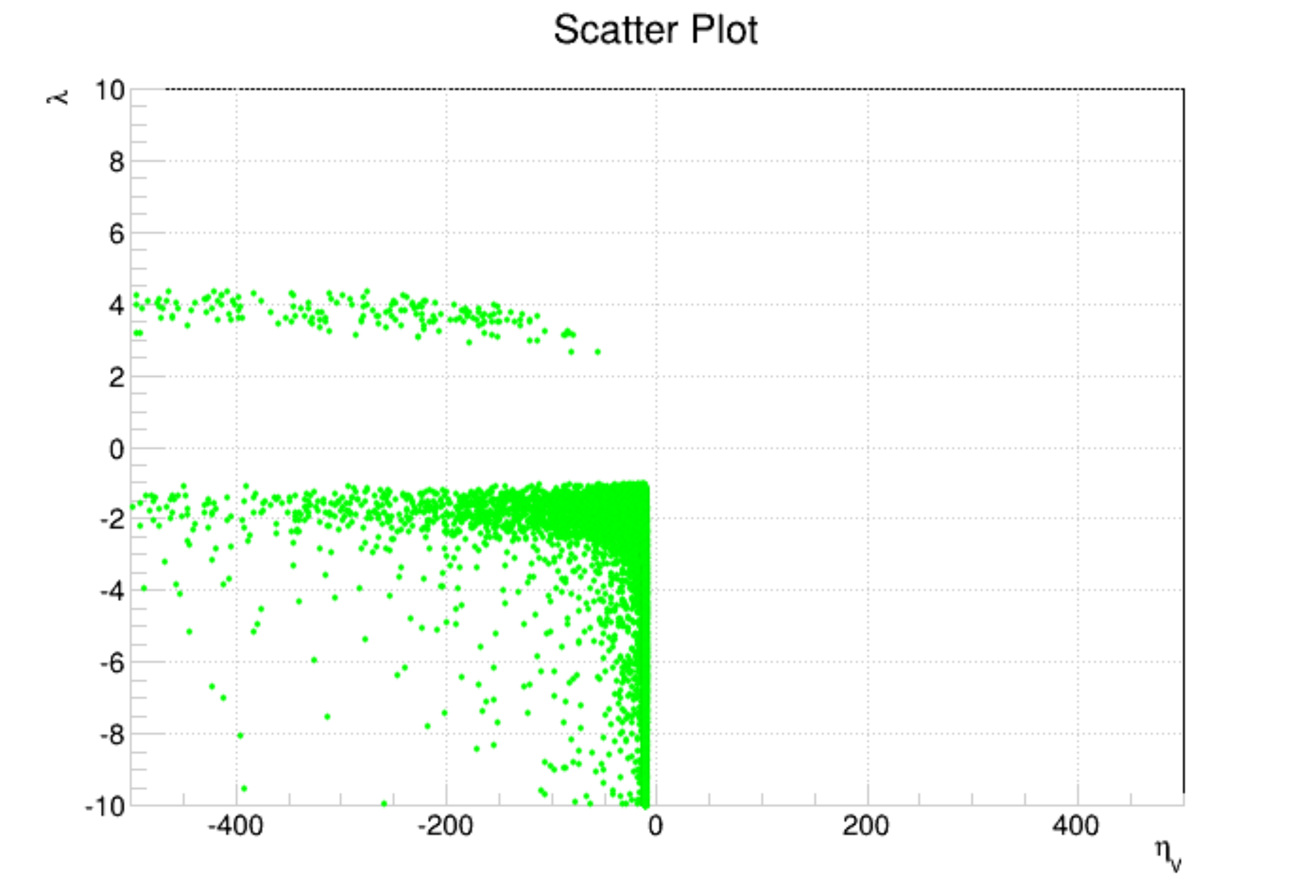}
  \captionof{figure}{Scatter plot of solution space showing the bound ``wall" close to a $-{\cal O}(1)$ value. }
  \label{fig:test1}
\end{minipage}%
\end{figure}
Now the left-hand sides of these equations can also be described directly in terms of the basic fields, $\rho, \tau, \sigma$ and $b$. We call these alternatively defined (but identical) quantities as
\bea
V_0^{'}=\frac{\rho^3}{\tau^4}, \ \
V_2^{(1^{'})}=\frac{3 b^2\rho}{\tau^4}, \ \ 
V_2^{(2^{'})}=\frac{3\rho}{\tau^4}, \ \
V_H^{(2^{'})}=\frac{9b^2}{\rho^3\tau^2\sigma^9}
\eea
We then define  
\bea
x_0= |\frac{V_0-V_0^{'}}{V_0+V_0^{'}}|, ~~x_{21}= |\frac{V_{2}^{(1)}-V_{2}^{(1^{'})}}{V_{2}^{(1)}+V_{2}^{(1^{'})}}|,
 ~~x_{22}= |\frac{V_{2}^{(2)}-V_{2}^{(2^{'})}}{V_{2}^{(2)}+V_{2}^{(2^{'})}}|,~~x_{H2}= |\frac{V_{H}^{(2)}-V_{H}^{(2^{'})}}{V_{H}^{(2)}+V_{H}^{(2^{'})}}| 
 \eea
To solve for the critical value, we scan the parameter space of 
$\lambda \in [-10, 10]$, $\rho, b \in [0, 2.5]$, $\sigma \in [0, 1.5]$ and $\tau \in [0, 0.5]$ while demanding $V_0, V_{2}^{(1)},V_{2}^{(2)},V_{H}^{(2)} >0 $ and 
\bea
x_0,x_{21},x_{22},x_{H2} \leq \Delta \equiv 0.05. \label{one-param}
\eea
Note that our resolution for calling a solution a solution is not too high, but this is enough to establish that these (potential) solutions satisfy the bound we wish to establish. In particular, note that these conditions on the $x$'s is merely a {\em necessary} condition for the existence of critical points: it is only in the limit that the resolution $\Delta$ goes to zero, that it becomes identical to the criticality condition, elsewhere, it may not be close to it. But we find  that these necessary conditions are quite useful: if $\Delta$ is too large the bound is not satisfied, but as we make the value of $\Delta$ smaller and smaller, a bound slowly emerges. the plot we present here is for the value $\Delta =0.05$.  The correct way to think about our approach is as a specific one-parameter ($\Delta$) generalization \eqref{one-param} of the critical point condition: the striking thing is that we find evidence for the bound-wall even in this weaker context.

\begin{figure}
\centering
\begin{minipage}{.8\textwidth}
  \centering
  \includegraphics[width=\linewidth]{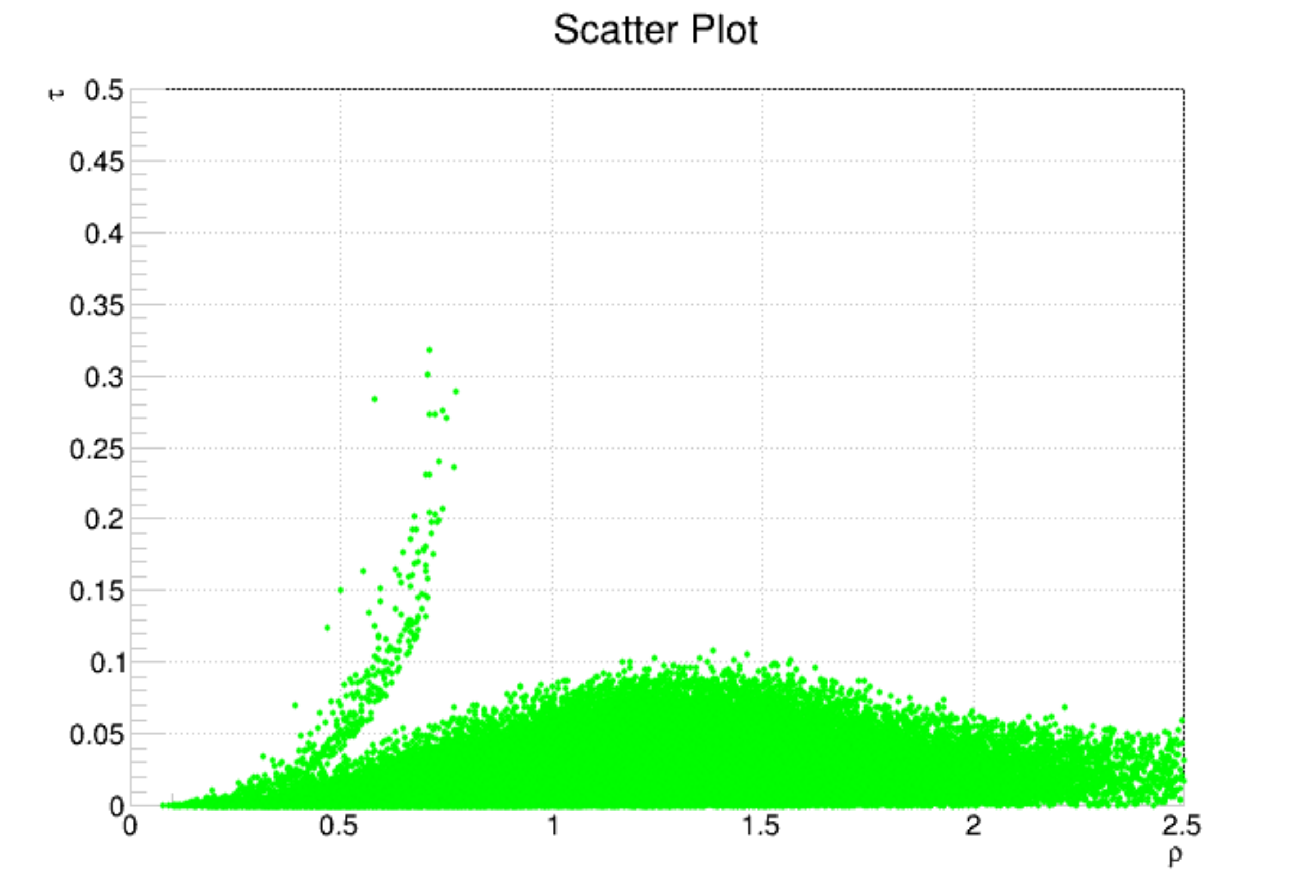}
  \captionof{figure}{Solution space in the $\tau-\rho$ directions.}
  \label{fig:test2}
\end{minipage}
\begin{minipage}{.8\textwidth}
  \centering
  \includegraphics[width=\linewidth]{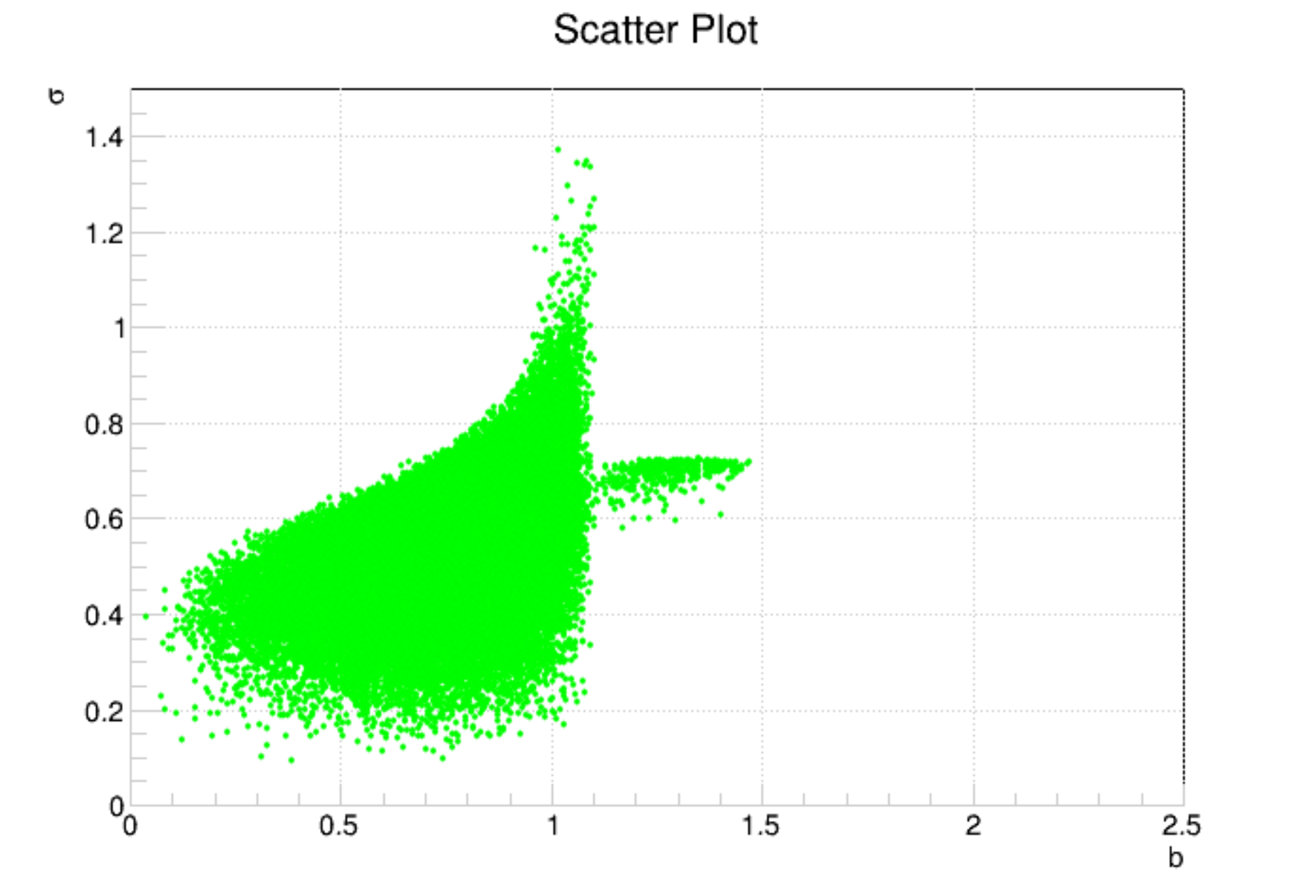}
  \captionof{figure}{Solution space in the $\sigma-b$ directions.}
  \label{fig:test2}
\end{minipage}
\end{figure}

Once we find these candidate solutions, we evaluate the minimum eigenvalues of the second derivative matrix of the potential. To do this without much hassle in the scan, it is somewhat useful to express the second derivatives in terms of the potential pieces themselves using the homogeneity properties of these pieces, and we collect the result in an Appendix. Note that at a critical point, the inhomogeneous piece in the covariant derivatives do not contribute, so we can self-consistently drop them since we are only looking for necessary conditions for the existence of the bound in this approach. The net result is that we get a scatter plot of the (possible) solutions, which we plot in the figure. In the plot, the $x$-axis is the $\eta_V$, the $y$-axis is unimportant, and it is evident that the solutions do not cross a ``wall" to the left of the origin. This happens at\footnote{We have checked that decreasing $\Delta$ does not change this value, even though it becomes significantly harder for us to find candidate solutions with smaller $\Delta$ with our strategy/resources.} about $\eta_V\approx-10$. The value could change by a numerical factor of ${\cal O}(1)$, if we were more careful with the pre-factors in \eqref{scalign} which lead to canonical kinetic terms: but since we worked with the reduced potential, it is not particularly meaningful to do so\footnote{We have checked that the magnitude of the bound becomes numerically slightly smaller (but still an ${\cal O}(1)$ number) because of these pre-factors in cases that only involve the universal moduli \cite{GK}.}. Overall, these facts are a strong indication that the ${\cal O}(1)$ bound indeed holds here as well, at least for generic tachyons. We also present plots of the ranges of field values for which we have searched for solutions. For $b, \tau, \sigma$ the plots suggest that we have exhausted --to within our numerical limitations-- candidate solutions in the (fairly small) field values that we consider. This does not rule out the existence of solutions isolated from these at larger field values. Even though our scan of $\rho$ is incomplete, we have checked that the qualitative behavior of the bound has stabilized in the range where we have scanned. Of course, a much more exhaustive scan needs to be done before a complete statement can be made, but we believe the ``wall'' we find is a good reason to think that most tachyons satisfy the ${\cal O}(1)$ bound.

\section{A Generalized Slow Roll Bound}

Our discussion in this section will be in the context of non-supersymmetric solutions with positive potential energy, because supersymmetric solutions are qualitatively different in many ways (they are necessarily stable, they have continuous moduli spaces, etc.). Our starting observation is the ``experimental'' fact in tree level string constructions that it has been impossible to construct de Sitter {\em vacua,} and that whenever de Sitter {\em solutions} exist, they have steep (ie., ${\cal O}(1)$ in appropriate Planck units) unstable directions. 

According to the bound in \cite{Vafa1}, on a generic point in the landscape, we expect the magnitude of the gradient of the potential to be $\gtrsim {\cal O}(1)$. Note that the $\epsilon_V$ parameter is defined in terms of sums of squares of the derivatives of the potential, so this means that the smallest possible slope in any direction is attained by distributing the slopes equally in all directions\footnote{Our goal is to {\em bound} slow roll, so we are interested in the minimum slope possible along {\em any} path at a point. Note also that as suggested in \cite{Achucarro-Palma,GK}, having a large number of fields is a way to make these bounds less constraining.}. This suggests that at a generic point in the landscape, there should exist a path along which one can flow so that the potential decreases by a fraction $\gtrsim {\cal O}(1)/\sqrt{N}$ as one moves by a field distance of about ${\cal O}(1)$, raising the possibility that one can make an integrated version of the slow roll bound statement. Here $N$ is the dimensionality of the field landscape. The argument above however is a bit naive because in principle, it is conceivable that the path hits a non-generic (ie., critical) point from which there is no way but up\footnote{Note that this is the {\em only} scenario in which one can {\em not} have a downward directed path from a point.}. But precisely at this stage, the slow roll bound of \cite{GK} comes to our rescue. It states that at any critical point, there exists a direction along which the second derivative is downward, and $\gtrsim {\cal O}(1)$ in Planck units. The fact that the magnitude of the second derivative is necessarily large suggests that the first derivative quickly re-approaches its generic value, which is bounded by $\sim {\cal O}(1)/\sqrt{N}$. 
Altogether, this sequence of arguments is one way to motivate that 
there exists a path of decreasing potential along which the following relations hold:
\bea
\ln\frac{V(\phi_1)}{V(\phi_0)}=\int \frac{\nabla_I V(\phi)}{V(\phi)} \frac{d \phi^I (s)}{ds} \, ds \lesssim  -\frac{ {\cal O}(1)}{\sqrt{N}} \int \left|\frac{d \phi^I (s)}{ds}\right|  ds = -\frac{ {\cal O}(1)}{\sqrt{N}} \int ds\label{argument}
\eea
%\int \left|\frac{\nabla_I V(\phi)}{V(\phi)}\right|\left|\frac{d \phi^I (s)}{ds}\right| ds\lesssim 
The first step is just the fundamental theorem of calculus written for a line integral of a gradient. The second inequality is pointwise true for the integrand on the path, if we choose the path correctly (there can be a few critical points where it might not be true, but they are a negligible contribution was the point of the discussion before). The norm is defined in the (Kahler) metric on the field space, and if we choose $s$ to be the arc length, that means that $\left|\frac{d \phi^I (s)}{ds}\right|$ is just unity. Note that the last expression involves the path length from $\phi_0$ to $\phi_1$. Therefore, the existence of a monotonically decreasing path to $\phi^I(s)$ from $\phi^I(0)$ where 
\bea
\ln\frac{V(\phi(s))}{V(\phi(0))} \lesssim -\frac{ {\cal O}(1)}{\sqrt{N}} s \label{general}
\eea
holds\footnote{Note that the only place where we have made an assumption in getting here is in the second step of \eqref{argument} where we have assumed that the inequality is sufficiently reasonable, thanks to the slow roll bound.}, (here $s$ is the path length) for {\em any}  $\phi_0$ is what emerges as a natural definition of a bound on slow roll. %The bound is tightest for the shortest such path. 

The above inequality is the most general form of the bound, but for practical purposes a convenient field range that one can consider is when the path length\footnote{Even though from the technical argument above, it is clear that the relevant length is the path length, in our examples they end up being identical to straightforward field ranges.} is $\lesssim {\cal O}(1)$. This is natural for a few different reasons, including the Field Range Bound \cite{FDB}. Not also  that we want to give enough room for the integrals to stabilize, and not be affected by the approximation we made while ignoring the contributions from near critical or non-generic points. All of this leads us to the most practically useful form that we will use for the {\em generalized slow roll bound}: 

%like in $N$-flation \cite{Nflation} 
%_{\phi^I_0}^{\phi^I_1} 

A natural notion of a slow roll bound in a multi-dimensional landscape is that for any point $\phi_0$ in field space with $V(\phi_0) > 0$, there exists a path (of monotonically decreasing\footnote{This path may in principle include critical points, but we expect that they can be made to avoid them by small perturbations that still satisfy the conditions of the bound. Note also that we can slightly relax the bound even further by demanding only monotonically non-increasing as opposed to decreasing. But this would mean that we are trying to include moduli directions at positive potential energy, which might not be particularly meaningful in a tree level construction. Such a construction has come  up after v1 of this paper appeared by including KK monopole sources \cite{Dib2}. We thank G. Dibitetto for a correspondence on this matter. This example violates all known dS swampland bounds including the present one, but because of the modulus direction it is not trustable as a purely tree-level solution.  It will be very interesting to see what happens to this solution if we can give a potential for this modulus using (reliable) quantum corrections. That will give a very strong hint for deciding whether our bounds should be viewed as bounds that should hold everywhere on the landscape or only at the boundary. 
%Even though there is a modulus at the critical point there and therefore its status is somewhat unclear, it will be interesting to see if the ``monotonically non-increasing'' version of our current bound holds in that case. 
} potential energy) to a point $\phi_1$ within a field distance ${\cal O}(1)$, such that  
\bea
\sqrt{N} \ln \frac{V(\phi_1)}{V(\phi_0)} \lesssim - {\cal O} (1). \label{BigBound}
\eea 
Note that if the potential hits zero within an ${\cal O}(1)$ field distance along the path, the left hand side blows up and the bound is trivially satisfied. It can also be easily checked that in the single field case, the previous slow roll bound \cite{GK} in fact implies this refined bound, and that exponential potentials saturate the bound. Note also that for the single field case at most the potential can have one maximum, because of the bound, and is otherwise monotonic. %The rest of the critical points, if any, are unstable.

We can view the new bound loosely as an integrated (non-local in field space) version of the slow roll bound, but it is more general than the previous versions for at least two reasons.

Firstly, even though we used the condition on the second derivative to argue that the path must go down when it hits a critical point, this can be relaxed while still retaining the above bound. All that is really required is that the higher derivatives at the critical point conspire in such a way that there is a downward path emerging from it, and that the derivatives controlling that direction are ${\cal O}(1)$ so that the first derivative re-approaches the generic value reasonably fast. For example, it is in principle possible that the second derivatives are zero at a critical point, and the higher derivatives are what ensure the downward flow of the trajectory. We present a simple example potential in an appendix that captures some of these features. Of course, as it stands critical points are already rare, so finding one where some of the second derivatives are also zero is likely to be extremely rare\footnote{See closely related results in \cite{wasteland}, where it was argued that merely having {\em all} positive eigenvalues in a 4D supergravity set up is statistically {\em highly} unlikely. Having a set of eigenvalues that are all zero, would presumably be measure zero in some suitable sense.}, so it is unsurprising that the such points have never been found (if at all they exist).  One can also come up with similar other examples as well. %Note that even finding the critical points and establishing their stability properties is highly non-trivial.

A second point that makes our new bound more general, is that  the original slow roll bounds are blind to points in the landscape that are ``close'' to a critical point, but not ``too close''. In the vicinity of critical points, by continuity, we might expect values of $\epsilon_V$ and 
$\eta_V$ that are in an intermediate range (instead of being hierarchically large or small). The current versions of the slow roll bound are not capable of handling these points, but the above version, is. Indeed, in the next section we will present an example from a known Type IIA string compactification which fits into this scenario. This example will illustrate the generality of the present version of the slow roll bound.

\subsection{A Hierarchy of Naturalness Bounds}

The non-local bound we have presented fits naturally into a hierarchy of bounds suggested by naturalness. To motivate it, first note that \cite{Vafa1} the cosmological ``constant'' problem can be viewed as the desire to fine tune the potential energy $V$ to hierarchically small values, and naturalness as the claim that its natural value is ${\cal O}(1)$ in Planck units. Now, let us assume that somehow we manage to get past this difficulty, and have managed to realize a potential energy $V$ that is hierarchically small. The claim of \cite{Vafa1} can then be understood as the statement that the natural value of its derivative (divided by the potential), is again ${\cal O}(1)$ in Planck units\footnote{Since this matter has come up in some correspondences that we have had, let us to emphasize that the \cite{Vafa1, GK} claim is {\em stronger} than just de Sitter No Go theorems. It is really a type of naturalness argument.}. Now, we know from ``experiment'' (aka tree level string constructions) that this bound is violated, and that in fact even this quantity can be hierarchically small. That is where the slow roll bounds of \cite{GK} kick in. It bounds the second derivative, when the first derivative is zero (or close to zero\footnote{Note that as long as the potential is continuous and the second derivative is non-zero, values of the first derivative close to zero should be allowed, if zero (critical point) is allowed.}). Notice the hierarchy of bounds: the original swampland criterion of \cite{Vafa1} argued that the first derivative of the potential should naturally be controlled by the Planck scale. Our refinement to that in \cite{GK} suggested instead that the first derivative can be zero, as long as the second derivative is controlled by the Planck scale\footnote{Of course, while forbidding de Sitter vacua.}. In other words, the claim of \cite{GK} can be thought of as the statement that the natural value of a non-zero derivative of the potential (divided by the potential itself) is ${\cal O}(1)$ in Planck units. This will basically ensure that the only fine-tuning in the problem is that of the vacuum energy, the derivatives when they are non-zero are natural, given the value of the vacuum energy.

The claims of \cite{Vafa1, GK} are both statements about the {\em potential} and do not concern themselves with the equations of motion. Therefore if they are true, it is a bit surprising why these conditions have to do with only the first two derivatives\footnote{In particular, note that for tree level constructions, there is no real reason even to think that the equations of motion have to necessarily be second order: tree level sigma model corrections to supergravity can give rise to higher order equations of motion. Such scenarios have not been investigated too much in the literature on de Sitter compactifications, as far as we know, but it is a logical possibility.}. A natural conjecture therefore   is to drop the demand that we are only dealing with the first two derivatives, and simply demand that  the non-zero derivatives are naturally controlled by the Planck scale. Our generalized slow roll bound has room to naturally incorporate this while still disallowing de Sitter vacua, as we mentioned earlier (see also the Appendix B). 
Our bound is a tightening of the crude idea that there exists paths in the landscape along which if one moves and ${\cal O}(1)$ distance, one should expect a drop by an ${\cal O}(1)$ fraction in the potential (upto subtleties associated with number of fields). 

The idea that higher order slow roll parameters can be defined and used is not new, and has been explored before, especially in the context of single field slow roll \cite{Liddle-Barrow}. Let us re-emphasize here that our interest here is in the so-called potential slow roll parameters, which only captures the shape of the potential. They do not say anything about the inflationary initial conditions and require more data (an ``attractor constraint" in single field slow roll) to fully fix inflationary dynamics. 

Note that since the claim on the first derivative in \cite{Vafa1} holds generically in large classes of tree level string constructions, it automatically implies that our present conjecture also holds generically. In fact, all known explicit tree level string constructions\footnote{Except for a caveat that we will get to momentarily.} (including large classes of non-explicit ones) that we are aware of satisfy either the \cite{Vafa1} bound or the bound on the second derivative in \cite{GK}. It is also easy to make explicit (but obviously non-exhaustive) checks of this in potential landscapes that descent from string theory. 

%\footnote{The statement of \cite{Vafa1} can be viewed as the statement that ``half of" the curves from a generic point satisfy \eqref{general} for small enough $s$.}.

%Since the present bound is slightly weaker than the previous slow roll bounds, it follows that it also holds in these set ups. It is straightforward to check the 

We will conclude this section with some comments about higher dimensional origins of these new bounds. The original dS swampland bounds \cite{HKTT, Vafa1}, and dS No-Go claims in general, have close connections to energy conditions in higher dimensions. So it is natural to think that our present bounds should also have an understanding in terms of them. We will have more to say about this in an upcoming paper \cite{future}. Here we will merely note that the energy conditions which are believed to be valid with quite some generality are the Averaged Null Energy Condition (ANEC): in Minkowski space, this is known to be derivable from the monotonicity of relative entropy \cite{Faulkner}.

\section{Application: An Accelerated Cosmology in Type IIA}

There exists one example that we are aware of in a tree level string construction where the bounds in \cite{Vafa1, GK} are under some tension. This corresponds to a class of torus orientifold compactifications in type IIA, where using a genetic algorithm based search, a few points on the tree level landscape were identified \cite{Dibitetto} with $\epsilon_V$ and $\eta_V$ both of ${\cal O}(0.1)$. In this scenario, the Universe undergoes a few e-folds, but not too many. In principle this is not a serious cause for concern since the slow roll bounds are order-of-magnitude bounds, and they are only violated within a factor of $\sim 10$ by these examples. Nonetheless, there is a conceptual problem that they bring to focus: the slow roll bounds only deal with the situation where {\em either} $\epsilon_V \gtrsim {\cal O}(1)$ and the bound holds as suggested in \cite{Vafa1}, {\em or}  $\epsilon_V \ll {\cal O}(1)$ and the bound holds as in \cite{GK}. But both bounds are silent about the region where $\epsilon_V$ (and $\eta_V$) is intermediate-valued. This is unsatisfactory: since the potential and its derivatives are continuous, near a critical point, we expect $\epsilon_V$ to sweep out intermediate values as well. Indeed, that this happens in tree-level string theory is what the results of \cite{Dibitetto} emphasize. 

Before we check our bound, lets set some expectations via crude estimates. We can start with a Taylor expansion around a point that is close to (but not identical to) a critical point up to second derivatives, perhaps with appropriate ${\cal O}(1)$ (in Planck units) coefficients. In such a scenario, very roughly, an ${\cal O}(1)$ combination of $\sqrt{\epsilon_V}$ and $\eta_V$ is what we really would expect to get bounded. Note that if both quantities are ${\cal O}(0.1)$ as in \cite{Dibitetto}, with appropriate ${\cal O}(1)$ coefficients, the resultant quantity could very well turn out to be bounded at $\gtrsim {\cal O}(1)$. Armed with this plausibility argument, now we can go ahead and check whether the potentials described in \cite{Dibitetto} satisfy our generalized slow roll bound around the field points corresponding to the initial conditions they use.  

What we will see is that while there are usually many directions along which the ${\cal O}(1)$ fall happens at a typical point in the landscape, the points found by \cite{Dibitetto} are atypical: if we are not aiming at those specific directions, we will be left with the impression that one is surrounded by peaks and there is ``no sufficiently steep way down''. However, we will also clearly see that once one finds the way down, it indeed satisfies our generalized bound on slow roll in this paper, namely that within an ${\cal O}(1)$ distance in field space, the potential drops by (more than) about an e-fold, upto the the $\sqrt{N}$ factor. We also suspect that there are (unstable) de Sitter extrema in these landscapes, but we will not try to establish that here\footnote{That de Sitter extrema do exist in the isotropic slice of the parameter space is known, see Appendix of the PhD thesis  \cite{PhD} for a discussion in the STU language close to ours.}.

\subsection{The ${\cal N}=1, D=4$ Effective Potential} 

The authors of \cite{Dibitetto} consider type IIA compactifications with O$6$/D$6$ sources on twisted tori of the form $T^{6}/(\mathbb{Z}_{2}\,\times\,\mathbb{Z}_{2})$. The details are unimportant for our purposes, what matters is that their effective four dimensional supergravity theory is an STU model whose effective potential we can calculate in terms of seven complex fields
$\Phi^{\alpha}\,\equiv\,\left(S,T_i,U_i\right)$ with $i=1,2,3$ as
\bea
\label{V}
V = e^{K}\left(-3\ |W|^{2}\ +\ K^{\alpha\bar{\beta}} \ D_{\alpha} W\ D_{\bar{\beta}}\overline{W} \right)  ,
\eea
with the Kahler potential 
\bea
K = -\log\left(-i (S-\overline{S})\right) - \sum_{i=1}^{3}{\log\left(-i (T_{i}-\overline{T}_{i})\right)} - \sum_{i=1}^{3}{\log\left(-i (U_{i}-\overline{U}_{i})\right)}\ .\label{Kahler}
\eea
and superpotential
\bea
\label{superpot}
W =P_{1}(U_{i})+S\,P_{2}(U_{i}) +\sum\limits_{k}{T_{k}\,P_{3}^{(k)}(U_{i})\,}
\eea
where $P_{1}$, $P_{2}$ and $P_{3}^{(k)}$ are (all indices run from 1 to 3, and parentheses do not indicate symmetrization)
\bea
\begin{array}{cclc}
P_1(U_i) & = & a_0\,-\sum\limits_i{a_1^{(i)} U_i} + \sum\limits_i{a_2^{(i)} \dfrac{U_1 U_2 U_3}{U_i}} - a_{3}U_1U_2 U_3 &  \\[3mm]
P_2(U_i) & = & -b_0+\sum\limits_i{b_1^{(i)} U_i} &  \\[3mm]
P_3^{(k)}(U_i) & = & c_0^{(k)}+\sum\limits_i{c_1^{(ik)}U_i} & 
\end{array}
\eea
The $a, b, c$ coefficients are fixed by the 10 dimensional (geometric) fluxes, and what \cite{Dibitetto} did was to identify four sets of values of these fluxes so that the slow roll parameters $\epsilon_V$ and $\eta_V$ are about ${\cal O}(0.1)$, and where the Universe undergoes a few e-folds. The Table of these flux values can be found in Appendix A of \cite{Dibitetto}, we will not repeat it. We quote the values of the corresponding slow roll parameters in a table below. They launch their evolutions from the origin of field space, so that is what we will adopt as well. 

\subsection{Checking the Bound}

\begin{figure}
\centering
\begin{minipage}{.7\textwidth}
  \centering
  \includegraphics[width=1.1\linewidth]{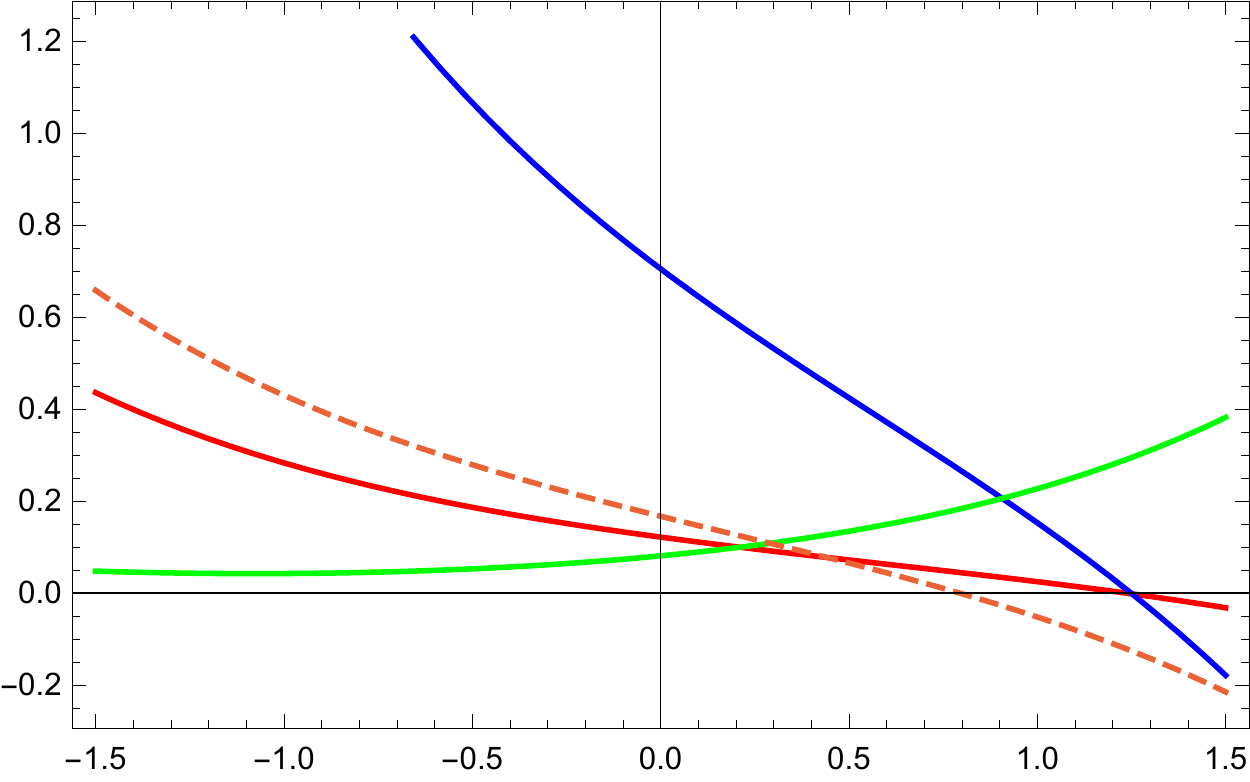}
  \captionof{figure}{Plots of slices of the potentials for four different ``generic'' flux sets that illustrate the bound.}
  \label{figplot}
\end{minipage}%
\end{figure}
To check that the bound is satisfied, what we need to do is evaluate the scalar potential in terms of the real fields
\bea
\left\{\begin{array}{lclc}
S & = & \chi  +  i  e^{-\phi} & \\[1mm]
T_{i} & = & \chi^{(1)}_i  +  i e^{-\phi^{(1)}_i} &  \\[1mm]
U_{i} & = & \chi^{(2)}_i +  i \, e^{-\phi^{(2)}_i} &  
\end{array}\right.
\eea
and see whether there are any field directions around the origin along which (a) the potential drops, and (b) satisfies \eqref{BigBound}, within an ${\cal O}(1)$ distance on field space. The Kahler metric in all the relevant directions we discuss is trivial, so the distance is just the field displacement divided by a factor of $\sqrt{2}$, but we will quote field values. 

In a generic landscape, we find that we can often satisfy the bound by moving along many of the (elementary) field directions. As a simple example of a generic landscape, we will take the cases presented in the tables of Appendix A of \cite{Dibitetto}, and simply replacing $b^{(ij)} \leftrightarrow b^{(ji)}$. Any generic choices of flux values will do, but we do this merely for concreteness. The results are presented in the table below, including the corresponding value of the $\epsilon_V$ parameter. We have also indicated the direction along which the field drops the fastest. The $-\infty$ means that the potential has hit zero at that point.  It is easily seen that our bound is satisfied, just as the original bound \cite{Vafa1} on $\epsilon_V$ is, as expected at a generic point.  We also present the plots of the potential along those field directions.

\begin{center}

  \begin{tabular}{|l | c | r|r| }

    \hline

    Sol & $\epsilon_{V}$ & $\phi_1$&$\sqrt{N}~$ln$( \frac{V(\phi_{1})}{V(\phi_{0})})$ \\ \hline

    1 & 8.82238 &  $\phi^{(1)}_{2} = 1$ & $ -5.93599$ \\ \hline

     2 & 8.52555 &  $\phi^{(1)}_{2} = 1$ & $  -5.71397$ \\ \hline

      3 & 5.29049 &  $\phi = -1$ & $ -2.40476$ \\ \hline

       4 & 9.52556 &  $\phi^{(1)}_{2} \approx 0.794$ & $ -\infty$ \\ \hline

  \end{tabular}

\end{center}

For the atypical flux values reported in the Appendix of \cite{Dibitetto} however, we find that {\em none} of the elementary field directions lead to the bound being satisfied convincingly, and in a couple of the four cases presented, the bound is violated flagrantly if one only considers those directions. However, we found that there do exist non-generic directions along which the bound is indeed satisfied. In all four cases, it turns out that moving along
\bea
\phi=\phi^{(1)}_{1}=\phi^{(1)}_{2}=\phi^{(1)}_{3} ( \equiv \phi^{'} )
\eea
leads to satisfying the bound convincingly. In the table below, we show the slow roll parameters corresponding to these solutions\footnote{We have re-calculated the $\epsilon_V$ parameters for confidence in trusting our calculation of the (quite complicated 14-dimensional) potential, but the $\eta_V$ parameters, we are merely quoting from \cite{Dibitetto}.}, and the values at various ${\cal O}(1)$ field values. It is evident that even though the previous slow roll bounds are under tension, the refined bound we presented in this paper is nicely satisfied. We also present the profile of the potentials along this directions in a plot\footnote{Note that in the generic case, many directions show a steep enough fall, but the curves are uninteresting. But here, the direction of steep fall is hard to find, but it is pretty when you find it.}, and in fact it turns out that all the different cases are fit by the same function\footnote{All fields that are not indicated in the argument of $V$ are understood to be zero.} 
\bea
V_0\equiv V \left(\phi=\phi^{(1)}_{1}=\phi^{(1)}_{2}=\phi^{(1)}_{3}\equiv \phi^{'}\right)= C_1 \, e^{2\phi^{'}} - C_2\, e^{3 \phi^{'}} + C_3\, e^{4\phi^{'}}.
\eea
The values of the $C_i$ are also presented in the table. Note that  $V_0$ has a pair of complex critical points, at the values of the $C_i$. This structure of the potential gives us a very nice understanding of what is actually going on in these bounded slow roll landscapes. It will be interesting to see if this can be used to identify an unstable de Sitter critical point of the full effective potential.

\begin{center}

  \begin{tabular}{|l | l | c || r|r||c|c|c| }

    \hline

    Sol. & $\epsilon_{V}$ &$\eta_{V}$& $\phi^{'}$&$\sqrt{N}$log$( \frac{V(\phi_{1})}{V(\phi_{0})})$ &$C_1$&       $C_2$& $C_3$ \\ \hline

    &            &             & $-0.5$ & $ -1.92583$ &       &       &\\ 

     1& $0.430424$ & $-0.151163$ &$-0.7$  &$-2.84467$   &$0.275$&$0.332$&$0.132$\\  

     &            &             &$-1.0$  &$-4.42050$   &       &       &\\\hline                                                                              

    &           &             & $-0.5$ & $ -2.12846$ &       &       &\\ 

    `2 & $0.452066$ & $-0.0776699$&$-0.7$  &$-3.07922$   &$1.272$&$1.516$&$0.621$\\  

     &            &             &$-1.0$  &$-4.68177$   &       &       &\\\hline

    &            &             & $-0.5$ & $-1.96684$  &       &       & \\ 

    3 & $0.383057$ & $-0.162635$ &$-0.7$  &$-2.88274$   &$0.205$&$0.251$&$0.102$\\  

     &            &             &$-1.0$  &$-4.45198$   &       &       &\\\hline 

   &            &             & $-0.5$ & $-2.02860$  &       &       &\\ 

     4 & $0.391704$ & $-0.318953$ &$-0.7$  &$-2.96456$   &$0.463$&$0.554$&$0.223$\\  

     &            &             &$-1.0$  &$-4.55520$   &       &       &\\\hline

  \end{tabular}

\end{center}

\begin{figure}
\centering
\begin{minipage}{.7\textwidth}
  \centering
  \includegraphics[width=1.1\linewidth]{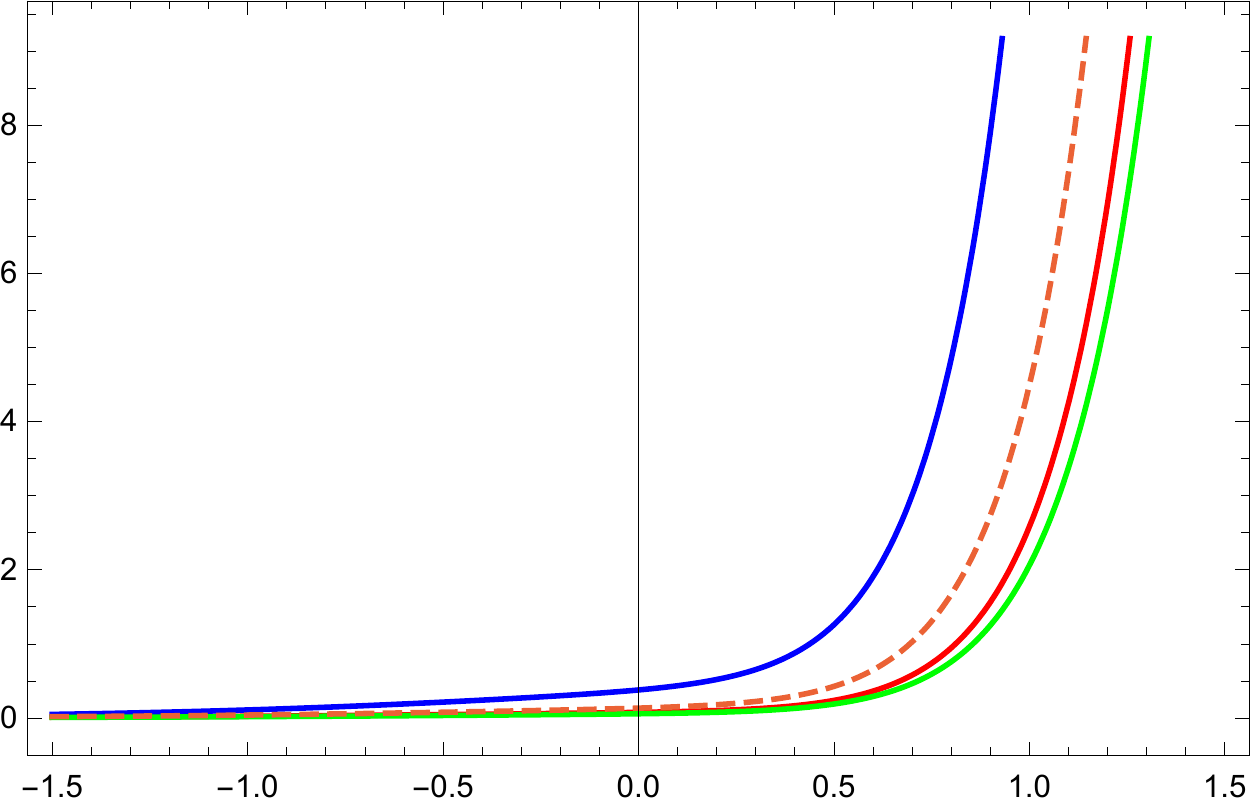}
  \captionof{figure}{Plots of slices of the potentials of the four different flux sets in \cite{Dibitetto} along the field direction that satisfies the bound.}
  \label{figplot}
\end{minipage}%
\end{figure}

\section{Outlook: Dine-Seiberg and Weak Coupling}

In this paper, instead of picking a decisive stance on the question of de Sitter vacua in string theory, we have stuck to more elementary questions which can be tackled more concretely at the present time. We hope this will ultimately contribute to a clear understanding of this issue and to forming a consensus. In particular, we believe it will be useful to precisely understand {\em when} the Swampland bounds and our generalizations of them hold. The claim of \cite{Vafa1, Vafa2} and other de Sitter denialists is that they hold everywhere in the string landscape. Our paper shows that appropriately refined versions of the Swampland bounds do hold quite generally in the context of {\em tree level} string/M-theory. This explains the title of our paper. At the same time, we are more skeptical about the full quantum validity of the bound.

We will conclude by suggesting that the swampland bounds are perhaps best thought of as a type of Dine-Seiberg argument. Note that if a potential is positive at large volume and goes to zero at infinite volume, then it will need a perturbative correction that is comparable to the tree level contribution, in order to introduce extra (hierarchically smaller) scales. This argument is very close to the standard one (say) in \cite{DenefLectures}, where it is used to conclude that existence of minima require perturbative corrections to be comparable to the tree level contribution. What we suggest, is that the same argument is also necessary for producing hierarchically smaller scales. Note that the bound in \cite{Vafa1}, and our re-interpretation of it as a statement about slow roll in \cite{GK} as well as in this paper, are both statements that there are no scales (in particular no hierarchically smaller scales) in the problem other than the Planck scale. Everything is ${\cal O}(1)$ in Planck units.

In other words, the statement that perturbation theory is arbitrarily well-defined, aka the statement that tree level is good enough, is the statement that if you move by ${\cal O}(1)$ in field space away from large radius, you should roughly expect an increase in your potential by a factor of ${\cal O}(1)$. Writing out this statement, and interpreting it in Taylor expansion, one ends up with various versions of the slow roll bounds. Of course, while we think this argument is quite suggestive, it can hardly be taken as proof. More work is certainly necessary to draw unambiguous conclusions about the regime of validity of the swampland bounds.

\section*{Acknowledgments}

We thank Giuseppe Dibitetto for a helpful comment in reproducing the ${\cal N}=1$ effective potential of arXiv:1310.8300 and K. V. Pavan Kumar for a useful discussion on landscape cartography. We also thank U. Danielsson, A. Hebecker, P. Hegde, A. Linde, S. Kachru, S. Trivedi, C. Vafa, T. Van Riet and T. Wrase for discussions/correspondence.

\appendix

\section{The Second Derivative Matrix for the Orientifold Tachyon}

The  elements of $4\times 4$ second derivative matrix can be re-expressed in terms of the homogeneity properties of the effective potential as,
\begin{eqnarray}
  M_{11} &=& \frac{\partial^2 V_{eff}}{\partial \hat{\rho}^2}=9 V_0+V_2+V_R+9V_H  \nonumber\\
 M_{22} &=& \frac{\partial^2 V_{eff}}{\partial \hat{\tau}^2}=16 V_0+ 16 V_2+9 V_{O6}+4 V_R+ 4 V_H   \nonumber\\
 M_{33} &=& \frac{\partial^2 V_{eff}}{\partial \hat{b}^2}=4 V_2^{(1)}-V_2^{(3)}+ 4 V_H^{(2)}-\lambda V_H^{(3)}+ 4 V_H^{(5)}+ \lambda V_H^{(6)}\nonumber\\
 M_{44} &=& \frac{\partial^2 V_{eff}}{\partial \hat{\sigma}^2}= \frac{81}{4}(\lambda-3)V_{O6}^{(1)}-\frac{9}{4}(\lambda+1) V_{O6}^{(2)}+81 V_R^{(1)}-9 V_R^{(2)}-9 V_R^{(3)}+81(\lambda^2 V_H^{(1)}+\nonumber\\&&\hspace{1.5cm}+V_H^{(2)}-\lambda V_H^{(3)})+9(\lambda^2 V_H^{(4)}+V_H^{(5)}+\lambda V_H^{(6)}) \nonumber\\
 M_{12} &=& \frac{\partial}{\partial \hat{\rho}}(\frac{\partial V_{eff}}{\partial \hat{\tau}}) =
 -12 V_0-4 V_2+ 2 V_R+ 6 V_H \nonumber\\
 M_{13} &=& \frac{\partial}{\partial \hat{\rho}}(\frac{\partial V_{eff}}{\partial \hat{b}})= 
 2 V_{2}^{(1)}-V_2^{(3)}-6 V_H^{(2)}+ 3\lambda V_H^{(3)}-6 V_H^{(5)}-3\lambda V_H^{(6)}\nonumber\\
 M_{14} &=& \frac{\partial}{\partial \hat{\rho}}(\frac{\partial V_{eff}}{\partial \hat{\sigma}})= 9 V_R^{(1)}-3 V_R^{(2)}+ 3 V_R^{(3)}+ 27(\lambda^2 V_H^{(1)}+V_H^{(2)}-\lambda V_H^{(3)})+\nonumber\\
&&\hspace{1.8cm} -9(\lambda^2 V_H^{(4)}+V_H^{(5)}+ \lambda V_H^{(6)}) \nonumber\\
 M_{23} &=& \frac{\partial}{\partial \hat{\tau}}(\frac{\partial V_{eff}}{\partial \hat{b}})= 
 -8 V_2^{(1)}+ 4 V_2^{(3)}-4 V_H^{(2)}+ 2\lambda V_H^{(3)}-4 V_H^{(5)}-2\lambda V_H^{(6)}\nonumber\\
 M_{24} &=& \frac{\partial}{\partial \hat{\tau}}(\frac{\partial V_{eff}}{\partial \hat{\sigma}})= \frac{27}{2}(\lambda-3) V_{O6}^{(1)}+\frac{9}{2}(\lambda+1) V_{O6}^{(2)}+ 18 V_R^{(1)}-6 V_R^{(2)}+6 V_R^{(3)}+\nonumber\\ && \hspace{1.8cm}+18(\lambda^2 V_H^{(1)}+V_H^{(2)}-\lambda V_H^{(3)})
-6(\lambda^2 V_H^{(4)}+V_H^{(5)}+\lambda V_H^{(6)})\nonumber\\
 M_{34} &=& \frac{\partial}{\partial \hat{b}}(\frac{\partial V_{eff}}{\partial \hat{\sigma}})= 
 -18 V_H^{(2)}+ 9\lambda V_H^{(3)}+ 6 V_H^{(5)}+ 3\lambda V_H^{(6)}\nonumber
\end{eqnarray}
The lowest eigenvalue of this matrix can be numerically evaluated using these.

\section{A Higher Order Example}

Let us present a small example to illustrate that vanishing first and second derivatives, but non-zero ${\cal O}(1)$ higher derivatives can still lead to bounds on the drop in the potential within an ${\cal O}(1)$ field range. Consider the example of the single field potential ($n>0$)
\bea
V(\phi) \sim 1 - \phi^n, \label{example}
\eea
around the origin, $\phi=0$. At the origin, the potential clearly has no minimum, but it is clear that it can in principle violate the bounds of \cite{Vafa1, GK} if $n>2$. However, such a potential does not violate a generalized slow roll bound, where the demand is merely that the drop in the potential should be at least by an ${\cal O}(1)$ fraction within an ${\cal O}(1)$ field distance. This is because the first non-zero derivative of the potential at the critical point is ${\cal O}(1)$ (with an appropriate sign that rules out de Sitter). 

Note in particular that even though it satisfies such a generalized bound, this sort of an inflationary potential can in principle allow sufficient e-foldings to explain horizon, flatness and related problems, around the origin of field space with appropriate initial conditions. This is analogous to the fact that in a multi-dimensional landscape, even when it satisfies the bounds of \cite{Vafa1, GK}, one can find inflationary trajectories that allow sufficient inflation \cite{Achucarro-Palma, GK}. This fact could be of significance in realizing inflation in tree level string landscapes, but such higher order critical points (if they exist at all) are likely to be exceedingly rare.

\begin{figure}
\centering
\begin{minipage}{.7\textwidth}
  \centering
  \includegraphics[width=1.1\linewidth]{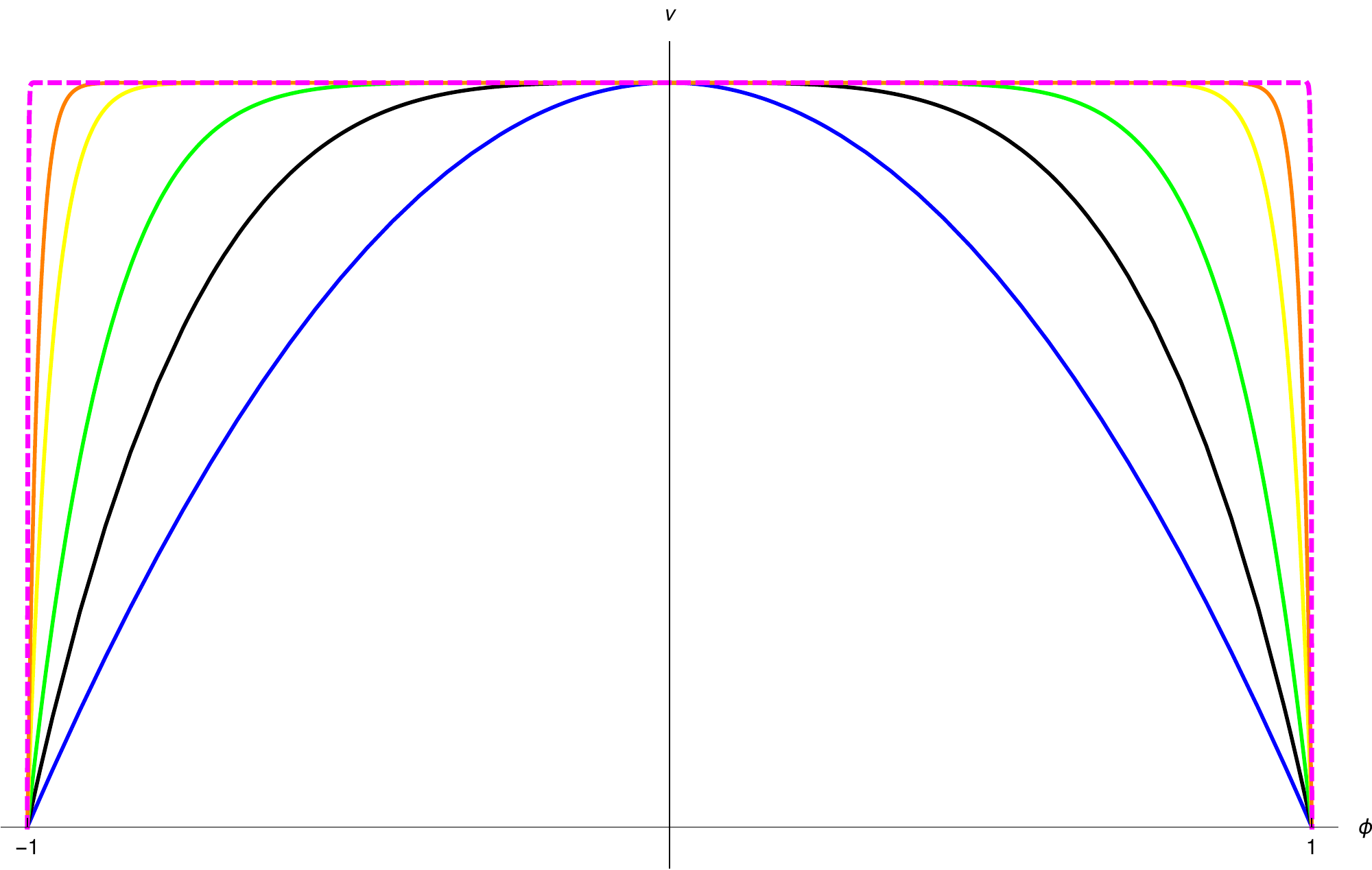}
  \captionof{figure}{Plot of the higher order potential for increasing values of (even) $n$. The point of interest to us here is that the value of the potential drops by an ${\cal O}(1)$ fraction within a field distance of ${\cal O}(1)$ for all values $n$.}
  \label{figplot}
\end{minipage}%
\end{figure}

\section{Cosmology: A Persistent Challenge for String Theory}

We have been fairly conservative in most of this paper, so let us take a moment to consider the most radical of all positions, 
more extreme than even the suggestions of \cite{Ulf, VafaReview, Sethi, Vafa1, Vafa2}: this is the possibility that perhaps string theory should be understood merely as a paradigm for understanding gauge theories holographically, and that it does not have anything to say about cosmology. String theory has had impressive successes in understanding various aspects of quantum gravity, black holes and gauge theories, and it contains the ingredients necessary for a successful phenomenology of particle physics. However, in its attempts at making sense of cosmology, there is a case to be made that it has been a failure at every step of the way -- or at the very least that the results from string theory have been nowhere  near as compelling in cosmology as it has been in these other arenas.

Lets discuss some of the pros and cons for believing that string theory does not have much to say about cosmology, and then we will take stock. Our conclusions are perhaps unsurprising: it seems to us that we have merely scratched the tip of the iceberg that is string theory, and so it is premature to conclude that it is not useful for cosmology despite its relative lackluster performance so far. 

\subsection{Pros}

\begin{itemize}
\item It is not clear how to quantize string theory in time dependent backgrounds (in particular cosmology). In fact things are worse: it is not clear how to quantize string theory in non-supersymmetric backgrounds in general.
\item String theory has had fairly good success in understanding singularities in time-{\bf in}dependent backgrounds, but time-like and null orbifold singularities even in flat space have been conceptually puzzling \cite{Seiberg, Horo}. Therefore the initial Big Bang singularity is a  mystery in string theory.
\item String theory seems extremely useful for understanding gauge theories holographically, especially when they have supersymmetry, but that success has not translated too effectively into a compelling holographic understanding of cosmology. 
\item The acceleration of the Universe is a challenge for string theory. Whether one views the acceleration as due to a cosmological constant or quintessence \cite{Willy, Susskind}, one runs into trouble with the fact that there is a cosmological horizon, and this leads to even more fundamental problems that will have to be addressed before even getting to string theory. (eg., what are the observables?). Note that a paper with a similar title to this section appeared 17 years ago. The de Sitter debate, still seems to be raging.
\item The multiverse/landscape solution to the accelerating Universe problem is distasteful to many.
\end{itemize}

\subsection{Cons}

\begin{itemize}
\item It is conceivable that the problem is more basic in some sense than the mere quantization of string theory in a time-dependent background, making this problem moot. Typically the quantization is done around a time-independent vacuum. A non-perturbative definition of string theory, as usual, will be very welcome. That would certainly make string theory more useful for cosmology. In fact it has not been possible to usefully quantize the string not just in time-dependent backgrounds, but even in flat space in the static gauge (but see \cite{Plefka}). The usual approach to (gauge-fixed) string quantization in flat space proceeds via the light-cone gauge. It is likely that quantization in the static gauge might reveal (at least) some technical insights required for an understanding of time-dependent backgrounds. So it is possible that at least some of these issues have also to do with technical difficulties. 
\item In an eternally inflating Universe, resolving the Big Bang singularity might be automatic. There has also been some progress in understanding Big Bang-like singularities in the tensionless limit of string theory \cite{CK-HS}.
\item dS/CFT correspondence is one attempt to describe cosmologies holographically \cite{Strominger}. It will be interesting to understand why the dS/CFT correspondence is possible. The holographic cosmology of Banks and Fischler \cite{Banks} is another set up for dealing with cosmology holographically. This has many promising features, but it is still in a fairly basic stage, and it will be interesting to understand its connection with string theory better. In all these situations, it is not clear how to make sense of quantum mechanics in a cosmological setting, to begin with. ``What are the observables?'', is a difficult question in any quantum mechanical setting of cosmology. This becomes especially vexing in a situation with cosmological horizons. This again, hints at the need for a more fundamental understanding of string theory before the problem can be addressed.
\item It is conceivable (even likely) that the landscape/multiverse is really how the Universe works. That the existence of numerous vacua is the only known solution to the cosmological constant problem should not be taken lightly.
\end{itemize}

To summarize: one of the primary difficulties is indeed that we are tied to supersymmetry for technical reasons for calculations. Also, it is perhaps premature to evaluate string theory's potential for dealing with cosmology, in the present limited context where it is still lacking a non-perturbative definition.  Furthermore, it will be very surprising indeed if we have a functioning quantum theory of gravity in AdS space, but our understanding of quantum gravity in a cosmological setting had nothing to do with that. Nature is afterall famous for picking the longest threads to weave its tapestry, to paraphrase a famous theorist. It seems quite possible that understanding the interior of the (small?) black hole in AdS might lead to progress in cosmology as well.


\begin{thebibliography}{99}

\bibitem{Carlos} 
  J.~M.~Maldacena and C.~Nunez,
  ``Supergravity description of field theories on curved manifolds and a no go theorem,''
  Int.\ J.\ Mod.\ Phys.\ A {\bf 16}, 822 (2001)
  doi:10.1142/S0217751X01003935, 10.1142/S0217751X01003937
  [hep-th/0007018].
  %%CITATION = doi:10.1142/S0217751X01003935, 10.1142/S0217751X01003937;%%
  %773 citations counted in INSPIRE as of 13 Jul 2018

\bibitem{Ulf} 
  U.~H.~Danielsson and T.~Van Riet,
  ``What if string theory has no de Sitter vacua?,''
  arXiv:1804.01120 [hep-th].
  %%CITATION = ARXIV:1804.01120;%%
  %8 citations counted in INSPIRE as of 13 Jul 2018

\bibitem{VafaReview} 
  T.~D.~Brennan, F.~Carta and C.~Vafa,
  ``The String Landscape, the Swampland, and the Missing Corner,''
  arXiv:1711.00864 [hep-th].
  %%CITATION = ARXIV:1711.00864;%%
  %24 citations counted in INSPIRE as of 13 Jul 2018

\bibitem{Sethi} 
  S.~Sethi,
  ``Supersymmetry Breaking by Fluxes,''
  arXiv:1709.03554 [hep-th].
  %%CITATION = ARXIV:1709.03554;%%
  %18 citations counted in INSPIRE as of 13 Jul 2018

\bibitem{many} 
  S.~B.~Giddings, S.~Kachru and J.~Polchinski,
  ``Hierarchies from fluxes in string compactifications,''
  Phys.\ Rev.\ D {\bf 66}, 106006 (2002)
  doi:10.1103/PhysRevD.66.106006
  [hep-th/0105097].
  %%CITATION = doi:10.1103/PhysRevD.66.106006;%%
  %1695 citations counted in INSPIRE as of 13 Jul 2018
 
   S.~Kachru, R.~Kallosh, A.~D.~Linde and S.~P.~Trivedi,
  ``De Sitter vacua in string theory,''
  Phys.\ Rev.\ D {\bf 68}, 046005 (2003)
  doi:10.1103/PhysRevD.68.046005
  [hep-th/0301240].
  

  
  %%CITATION = ARXIV:1509.05710;%%
  %27 citations counted in INSPIRE as of 16 Jul 2018
    V.~Balasubramanian, P.~Berglund, J.~P.~Conlon and F.~Quevedo,
  ``Systematics of moduli stabilisation in Calabi-Yau flux compactifications,''
  JHEP {\bf 0503}, 007 (2005)
  doi:10.1088/1126-6708/2005/03/007
  [hep-th/0502058].
  
  %%CITATION = doi:10.1088/1126-6708/2005/03/007;%%
  %671 citations counted in INSPIRE as of 16 Jul 2018
    A.~Westphal,
  ``de Sitter string vacua from Kahler uplifting,''
  JHEP {\bf 0703}, 102 (2007)
  doi:10.1088/1126-6708/2007/03/102
  [hep-th/0611332].
  
  %%CITATION = doi:10.1088/1126-6708/2007/03/102;%%
  %90 citations counted in INSPIRE as of 16 Jul 2018
  D.~Cohen-Maldonado, J.~Diaz, T.~van Riet and B.~Vercnocke,
  ``Observations on fluxes near anti-branes,''
  JHEP {\bf 1601}, 126 (2016)
  doi:10.1007/JHEP01(2016)126
  [arXiv:1507.01022 [hep-th]].
  

  E.~Silverstein,
  ``Simple de Sitter Solutions,''
  Phys.\ Rev.\ D {\bf 77}, 106006 (2008)
  doi:10.1103/PhysRevD.77.106006
  [arXiv:0712.1196 [hep-th]].
  %%CITATION = doi:10.1103/PhysRevD.77.106006;%%
  %136 citations counted in INSPIRE as of 22 Oct 2018  
  
  %%CITATION = doi:10.1007/JHEP01(2016)126;%%
  %21 citations counted in INSPIRE as of 16 Jul 2018
   I.~Bena, J.~Blåbäck and D.~Turton,
  ``Loop corrections to the antibrane potential,''
  JHEP {\bf 1607}, 132 (2016)
  doi:10.1007/JHEP07(2016)132
  [arXiv:1602.05959 [hep-th]].

  
  %%CITATION = doi:10.1007/JHEP03(2017)141;%%
  %7 citations counted in INSPIRE as of 16 Jul 2018
   M.~Bertolini, D.~Musso, I.~Papadimitriou and H.~Raj,
  ``A goldstino at the bottom of the cascade,''
  JHEP {\bf 1511}, 184 (2015)
  doi:10.1007/JHEP11(2015)184
  [arXiv:1509.03594 [hep-th]].
  
  %%CITATION = doi:10.1007/JHEP11(2015)184;%%
  %20 citations counted in INSPIRE as of 16 Jul 2018
  C.~Krishnan, H.~Raj and P.~N.~Bala Subramanian,
  ``On the KKLT Goldstino,''
  JHEP {\bf 1806}, 092 (2018)
  doi:10.1007/JHEP06(2018)092
  [arXiv:1803.04905 [hep-th]].
  
  %%CITATION = doi:10.1007/JHEP06(2018)092;%%
  %2 citations counted in INSPIRE as of 16 Jul 2018
   I.~Bena, M.~Grana and N.~Halmagyi,
  ``On the Existence of Meta-stable Vacua in Klebanov-Strassler,''
  JHEP {\bf 1009}, 087 (2010)
  doi:10.1007/JHEP09(2010)087
  [arXiv:0912.3519 [hep-th]].
  
  %%CITATION = doi:10.1007/JHEP09(2010)087;%%
  %128 citations counted in INSPIRE as of 16 Jul 2018
  E.~A.~Bergshoeff, K.~Dasgupta, R.~Kallosh, A.~Van Proeyen and T.~Wrase,
  ``$ \overline{\mathrm{D}3} $ and dS,''
  JHEP {\bf 1505}, 058 (2015)
  doi:10.1007/JHEP05(2015)058
  [arXiv:1502.07627 [hep-th]].
  %%CITATION = doi:10.1007/JHEP05(2015)058;%%
  %70 citations counted in INSPIRE as of 16 Jul 2018


  A.~Ashoorioon, H.~Firouzjahi and M.~M.~Sheikh-Jabbari,
  ``M-flation: Inflation From Matrix Valued Scalar Fields,''
  JCAP {\bf 0906}, 018 (2009)
  doi:10.1088/1475-7516/2009/06/018
  [arXiv:0903.1481 [hep-th]].
  %%CITATION = doi:10.1088/1475-7516/2009/06/018;%%
  %71 citations counted in INSPIRE as of 22 Oct 2018

K.~A.~Intriligator, N.~Seiberg and D.~Shih,
  ``Dynamical SUSY breaking in meta-stable vacua,''
  JHEP {\bf 0604}, 021 (2006)
  doi:10.1088/1126-6708/2006/04/021
  [hep-th/0602239].
  %%CITATION = doi:10.1088/1126-6708/2006/04/021;%%
  %598 citations counted in INSPIRE as of 16 Jul 2018

 W.~Fischler, V.~Kaplunovsky, C.~Krishnan, L.~Mannelli and M.~A.~C.~Torres,
  ``Meta-Stable Supersymmetry Breaking in a Cooling Universe,''
  JHEP {\bf 0703}, 107 (2007)
  doi:10.1088/1126-6708/2007/03/107
  [hep-th/0611018].
 
  D.~Junghans and M.~Zagermann,
  ``A Universal Tachyon in Nearly No-scale de Sitter Compactifications,''
  arXiv:1612.06847 [hep-th].
  %%CITATION = ARXIV:1612.06847;%%
  %3 citations counted in INSPIRE as of 13 Jul 2018

  G.~Dvali and C.~Gomez,
  ``On Exclusion of Positive Cosmological Constant,''
  arXiv:1806.10877 [hep-th].
  %%CITATION = ARXIV:1806.10877;%%
  %26 citations counted in INSPIRE as of 06 Nov 2018

  G.~Dvali, C.~Gomez and S.~Zell,
  ``Quantum Break-Time of de Sitter,''
  JCAP {\bf 1706}, 028 (2017)
  doi:10.1088/1475-7516/2017/06/028
  [arXiv:1701.08776 [hep-th]].
  %%CITATION = doi:10.1088/1475-7516/2017/06/028;%%
  %15 citations counted in INSPIRE as of 06 Nov 2018

  S.~Vagnozzi, S.~Dhawan, M.~Gerbino, K.~Freese, A.~Goobar and O.~Mena,
  ``Constraints on the sum of the neutrino masses in dynamical dark energy models with $w(z) \geq -1$ are tighter than those obtained in $\Lambda$CDM,''
  Phys.\ Rev.\ D {\bf 98}, no. 8, 083501 (2018)
  doi:10.1103/PhysRevD.98.083501
  [arXiv:1801.08553 [astro-ph.CO]].

%\cite{Lehners:2018vgi}
  J.~L.~Lehners,
  ``Small-Field and Scale-Free: Inflation and Ekpyrosis at their Extremes,''
  arXiv:1807.05240 [hep-th].

%\cite{Denef:2018etk}

%\cite{Colgain:2018wgk}
  E.~Ó.~Colgáin, M.~H.~P.~M.~Van Putten and H.~Yavartanoo,
  ``$H_0$ tension and the de Sitter Swampland,''
  arXiv:1807.07451 [hep-th].
%\cite{Roupec:2018mbn}
 
  C.~Roupec and T.~Wrase,
  ``de Sitter extrema and the swampland,''
  arXiv:1807.09538 [hep-th].
%\cite{Matsui:2018bsy}

  H.~Matsui and F.~Takahashi,
  ``Eternal Inflation and Swampland Conjectures,''
  arXiv:1807.11938 [hep-th].
%\cite{Ben-Dayan:2018mhe}

  I.~Ben-Dayan,
  ``Draining the Swampland,''
  arXiv:1808.01615 [hep-th].
%\cite{Damian:2018tlf}

  C.~Damian and O.~Loaiza-Brito,
  ``Two-field axion inflation and the swampland constraint in the flux-scaling scenario,''
  arXiv:1808.03397 [hep-th].
%\cite{Conlon:2018eyr}

  W.~H.~Kinney, S.~Vagnozzi and L.~Visinelli,
  ``The Zoo Plot Meets the Swampland: Mutual (In)Consistency of Single-Field Inflation, String Conjectures, and Cosmological Data,''
  arXiv:1808.06424 [astro-ph.CO].
%\cite{Dasgupta:2018rtp}

  K.~Dasgupta, M.~Emelin, E.~McDonough and R.~Tatar,
  ``Quantum Corrections and the de Sitter Swampland Conjecture,''
  arXiv:1808.07498 [hep-th]
%\cite{Marsh:2018kub}

  M.~C.~D.~Marsh,
  ``The Swampland, Quintessence and the Vacuum Energy,''
  arXiv:1809.00726 [hep-th].
%\cite{Brahma:2018hrd}

  S.~Brahma and M.~Wali Hossain,
  ``Avoiding the string swampland in single-field inflation: Excited initial states,''
  arXiv:1809.01277 [hep-th].
%\cite{Choi:2018rze}

  K.~Choi, D.~Chway and C.~S.~Shin,
  ``The dS swampland conjecture with the electroweak symmetry and QCD chiral symmetry breaking,''
  arXiv:1809.01475 [hep-th].
%\cite{Das:2018hqy}

  S.~Das,
  ``A note on Single-field Inflation and the Swampland Criteria,''
  arXiv:1809.03962 [hep-th].
%\cite{Han:2018yrk}

  C.~Han, S.~Pi and M.~Sasaki,
  ``Quintessence Saves Higgs Instability,''
  arXiv:1809.05507 [hep-ph].

%\cite{Visinelli:2018utg}

  L.~Visinelli and S.~Vagnozzi,
  ``A cosmological window onto the string axiverse and the supersymmetry breaking scale,''
  arXiv:1809.06382 [hep-ph].
%\cite{Moritz:2018ani}

  J.~Moritz, A.~Retolaza and A.~Westphal,
  ``On uplifts by warped anti-D3-branes,''
  arXiv:1809.06618 [hep-th].
%\cite{Matsui:2018xwa}

  H.~Matsui, F.~Takahashi and M.~Yamada,
  ``Isocurvature Perturbations of Dark Energy and Dark Matter from the Swampland Conjecture,''
  arXiv:1809.07286 [astro-ph.CO].
%\cite{Dimopoulos:2018upl}

  K.~Dimopoulos,
  ``Steep Eternal Inflation and the Swampland,''
  arXiv:1810.03438 [gr-qc].
%\cite{Anguelova:2018vyr}

  L.~Anguelova, E.~M.~Babalic and C.~I.~Lazaroiu,
  ``Two-field Cosmological $\alpha$-attractors with Noether Symmetry,''
  arXiv:1809.10563 [hep-th].
%\cite{Lin:2018kjm}

  C.~M.~Lin, K.~W.~Ng and K.~Cheung,
  ``Chaotic inflation on the brane and the Swampland Criteria,''
  arXiv:1810.01644 [hep-ph].
%\cite{Hamaguchi:2018vtv}

  K.~Hamaguchi, M.~Ibe and T.~Moroi,
  ``The swampland conjecture and the Higgs expectation value,''
  arXiv:1810.02095 [hep-th].
%\cite{Motaharfar:2018zyb}

  M.~Motaharfar, V.~Kamali and R.~O.~Ramos,
  ``Warm way out of the Swampland,''
  arXiv:1810.02816 [astro-ph.CO].

%\cite{Das:2018rpg}

  S.~Das,
  ``Warm Inflation in the light of Swampland Criteria,''
  arXiv:1810.05038 [hep-th].
%\cite{Ooguri:2018wrx}

  S.~J.~Wang,
  ``Quintessential Starobinsky inflation and swampland criteria,''
  arXiv:1810.06445 [hep-th].
%\cite{Fukuda:2018haz}

  H.~Fukuda, R.~Saito, S.~Shirai and M.~Yamazaki,
  ``Phenomenological Consequences of the Refined Swampland Conjecture,''
  arXiv:1810.06532 [hep-th].  

  A.~Hebecker and T.~Wrase,
  ``The asymptotic dS Swampland Conjecture - a simplified derivation and a potential loophole,''
  arXiv:1810.08182 [hep-th].
  %%CITATION = ARXIV:1810.08182;% 

A.~Kehagias and A.~Riotto,
  ``A note on Inflation and the Swampland,''
  arXiv:1807.05445 [hep-th].

  D.~Andriot,
  ``New constraints on classical de Sitter: flirting with the swampland,''
  arXiv:1807.09698 [hep-th].
  

 L.~Heisenberg, M.~Bartelmann, R.~Brandenberger and A.~Refregier,
  ``Dark Energy in the Swampland,''
  arXiv:1808.02877 [astro-ph.CO].
  
  Y.~Akrami, R.~Kallosh, A.~Linde and V.~Vardanyan,
  %``The Landscape, the Swampland and the Era of Precision Cosmology,''
  Fortsch.\ Phys.\  {\bf 67}, no. 1-2, 1800075 (2019)
  doi:10.1002/prop.201800075
  [arXiv:1808.09440 [hep-th]].
  %%CITATION = doi:10.1002/prop.201800075;%%
  %69 citations counted in INSPIRE as of 03 Apr 2019

  L.~Covi, M.~Gomez-Reino, C.~Gross, J.~Louis, G.~A.~Palma and C.~A.~Scrucca,
  ``Constraints on modular inflation in supergravity and string theory,''
  JHEP {\bf 0808}, 055 (2008)
  doi:10.1088/1126-6708/2008/08/055
  [arXiv:0805.3290 [hep-th]].


  A.~Hetz and G.~A.~Palma,
  ``Sound Speed of Primordial Fluctuations in Supergravity Inflation,''
  Phys.\ Rev.\ Lett.\  {\bf 117}, no. 10, 101301 (2016)
  doi:10.1103/PhysRevLett.117.101301
  [arXiv:1601.05457 [hep-th]].

  A.~Ashoorioon and M.~M.~Sheikh-Jabbari,
  ``Gauged M-flation, its UV sensitivity and Spectator Species,''
  JCAP {\bf 1106}, 014 (2011)
  doi:10.1088/1475-7516/2011/06/014
  [arXiv:1101.0048 [hep-th]].

  A.~Ashoorioon and A.~Krause,
  ``Power Spectrum and Signatures for Cascade Inflation,''
  hep-th/0607001.
  
 

  

\bibitem{Vafa1} 
  G.~Obied, H.~Ooguri, L.~Spodyneiko and C.~Vafa,
  ``De Sitter Space and the Swampland,''
  arXiv:1806.08362 [hep-th].
  %%CITATION = ARXIV:1806.08362;%%
  %7 citations counted in INSPIRE as of 13 Jul 2018

\bibitem{GK} 
  S.~K.~Garg and C.~Krishnan,
  ``Bounds on Slow Roll and the de Sitter Swampland,''
  arXiv:1807.05193 [hep-th].
  %%CITATION = ARXIV:1807.05193;%%
  %28 citations counted in INSPIRE as of 18 Oct 2018


  %%CITATION = doi:10.1088/1126-6708/2007/03/107;%%
  %82 citations counted in INSPIRE as of 16 Jul 2018



  %%CITATION = doi:10.1103/PhysRevD.68.046005;%%
  %2534 citations counted in INSPIRE as of 16 Jul 2018
 
\bibitem{Farhi-Guth} 
  E.~Farhi and A.~H.~Guth,
  ``An Obstacle to Creating a Universe in the Laboratory,''
  Phys.\ Lett.\ B {\bf 183}, 149 (1987).
  doi:10.1016/0370-2693(87)90429-1
  %%CITATION = doi:10.1016/0370-2693(87)90429-1;%%
  %167 citations counted in INSPIRE as of 15 Jul 2018 
 
\bibitem{Quigley} 
  C.~Quigley,
  ``Gaugino Condensation and the Cosmological Constant,''
  JHEP {\bf 1506}, 104 (2015)
  doi:10.1007/JHEP06(2015)104
  [arXiv:1504.00652 [hep-th]].
  %%CITATION = doi:10.1007/JHEP06(2015)104;%%
  %14 citations counted in INSPIRE as of 18 Oct 2018 
 
 \bibitem{Dine-Seiberg} 
  M.~Dine and N.~Seiberg,
  ``Is the Superstring Weakly Coupled?,''
  Phys.\ Lett.\  {\bf 162B}, 299 (1985).
  doi:10.1016/0370-2693(85)90927-X
  %%CITATION = doi:10.1016/0370-2693(85)90927-X;%%
  %283 citations counted in INSPIRE as of 18 Oct 2018
 
\bibitem{Polchinski}  
   J.~Polchinski,
  ``Brane/antibrane dynamics and KKLT stability,''
  arXiv:1509.05710 [hep-th].

%\cite{Agrawal:2018own}
\bibitem{Vafa2} 
  P.~Agrawal, G.~Obied, P.~J.~Steinhardt and C.~Vafa,
  ``On the Cosmological Implications of the String Swampland,''
  arXiv:1806.09718 [hep-th].
  %%CITATION = ARXIV:1806.09718;%%
  %5 citations counted in INSPIRE as of 13 Jul 2018

\bibitem{HKTT} 
  M.~P.~Hertzberg, S.~Kachru, W.~Taylor and M.~Tegmark,
  ``Inflationary Constraints on Type IIA String Theory,''
  JHEP {\bf 0712}, 095 (2007)
  doi:10.1088/1126-6708/2007/12/095
  [arXiv:0711.2512 [hep-th]].
  %%CITATION = doi:10.1088/1126-6708/2007/12/095;%%
  %126 citations counted in INSPIRE as of 13 Jul 2018


\bibitem{Andriot} 
  D.~Andriot,
  ``On the de Sitter swampland criterion,''
  arXiv:1806.10999 [hep-th].
  %%CITATION = ARXIV:1806.10999;%%
  %2 citations counted in INSPIRE as of 13 Jul 2018

\bibitem{Koerber} 
  U.~H.~Danielsson, S.~S.~Haque, P.~Koerber, G.~Shiu, T.~Van Riet and T.~Wrase,
  ``De Sitter hunting in a classical landscape,''
  Fortsch.\ Phys.\  {\bf 59}, 897 (2011)
  doi:10.1002/prop.201100047
  [arXiv:1103.4858 [hep-th]].
  %%CITATION = doi:10.1002/prop.201100047;%%
  %64 citations counted in INSPIRE as of 13 Jul 2018

\bibitem{Shiu} 
  G.~Shiu and Y.~Sumitomo,
  ``Stability Constraints on Classical de Sitter Vacua,''
  JHEP {\bf 1109}, 052 (2011)
  doi:10.1007/JHEP09(2011)052
  [arXiv:1107.2925 [hep-th]].
  %%CITATION = doi:10.1007/JHEP09(2011)052;%%
  %39 citations counted in INSPIRE as of 13 Jul 2018

\bibitem{VanRiet} 
  T.~Van Riet,
  ``On classical de Sitter solutions in higher dimensions,''
  Class.\ Quant.\ Grav.\  {\bf 29}, 055001 (2012)
  doi:10.1088/0264-9381/29/5/055001
  [arXiv:1111.3154 [hep-th]].
  %%CITATION = doi:10.1088/0264-9381/29/5/055001;%%
  %19 citations counted in INSPIRE as of 13 Jul 201


\bibitem{obstinate} 
  U.~H.~Danielsson, G.~Shiu, T.~Van Riet and T.~Wrase,
  ``A note on obstinate tachyons in classical dS solutions,''
  JHEP {\bf 1303}, 138 (2013)
  doi:10.1007/JHEP03(2013)138
  [arXiv:1212.5178 [hep-th]].
  %%CITATION = doi:10.1007/JHEP03(2013)138;%%
  %35 citations counted in INSPIRE as of 21 Oct 2018

\bibitem{Dibitetto} 
  J.~Blåbäck, U.~Danielsson and G.~Dibitetto,
  ``Accelerated Universes from type IIA Compactifications,''
  JCAP {\bf 1403}, 003 (2014)
  doi:10.1088/1475-7516/2014/03/003
  [arXiv:1310.8300 [hep-th]].
  %%CITATION = doi:10.1088/1475-7516/2014/03/003;%%
  %12 citations counted in INSPIRE as of 21 Oct 2018

\bibitem{Kallosh} 
  R.~Kallosh and T.~Wrase,
  ``dS Supergravity from 10d,''
  Fortsch.\ Phys.\  {\bf 2018}, 1800071
  doi:10.1002/prop.201800071
  [arXiv:1808.09427 [hep-th]].
  %%CITATION = doi:10.1002/prop.201800071;%%
  %4 citations counted in INSPIRE as of 21 Oct 2018

\bibitem{KPV} 
  S.~Kachru, J.~Pearson and H.~L.~Verlinde,
  ``Brane / flux annihilation and the string dual of a nonsupersymmetric field theory,''
  JHEP {\bf 0206}, 021 (2002)
  doi:10.1088/1126-6708/2002/06/021
  [hep-th/0112197].
  %%CITATION = doi:10.1088/1126-6708/2002/06/021;%%
  %333 citations counted in INSPIRE as of 21 Oct 2018

\bibitem{CK-Jarah} 
  J.~Evslin, C.~Krishnan and S.~Kuperstein,
  ``Cascading quivers from decaying D-branes,''
  JHEP {\bf 0708}, 020 (2007)
  doi:10.1088/1126-6708/2007/08/020
  [arXiv:0704.3484 [hep-th]].
  %%CITATION = doi:10.1088/1126-6708/2007/08/020;%%
  %12 citations counted in INSPIRE as of 21 Oct 2018

%\cite{Danielsson:2016cit}
\bibitem{Van} 
  U.~H.~Danielsson, F.~F.~Gautason and T.~Van Riet,
  ``Unstoppable brane-flux decay of $ \overline{\mathrm{D}6} $ branes,''
  JHEP {\bf 1703}, 141 (2017)
  doi:10.1007/JHEP03(2017)141
  [arXiv:1609.06529 [hep-th]].
  %%CITATION = doi:10.1007/JHEP03(2017)141;%%
  %10 citations counted in INSPIRE as of 21 Oct 2018

%\cite{Achucarro:2018vey}
\bibitem{Achucarro-Palma} 
  A.~Achucarro and G.~A.~Palma,
  ``The string swampland constraints require multi-field inflation,''
  arXiv:1807.04390 [hep-th].
  %%CITATION = ARXIV:1807.04390;%%1

\bibitem{FDB} 
  H.~Ooguri and C.~Vafa,
  ``On the Geometry of the String Landscape and the Swampland,''
  Nucl.\ Phys.\ B {\bf 766}, 21 (2007)
  doi:10.1016/j.nuclphysb.2006.10.033
  [hep-th/0605264].
  %%CITATION = doi:10.1016/j.nuclphysb.2006.10.033;%%
  %182 citations counted in INSPIRE as of 21 Oct 2018

%\bibitem{FDB2}   F.~Baume and E.~Palti,  ``Backreacted Axion Field Ranges in String Theory,''  JHEP {\bf 1608}, 043 (2016)  doi:10.1007/JHEP08(2016)043  [arXiv:1602.06517 [hep-th]].
  %%CITATION = doi:10.1007/JHEP08(2016)043;%%
  %40 citations counted in INSPIRE as of 21 Oct 2018

\bibitem{Dib2} 
  J.~Blåbäck, U.~Danielsson and G.~Dibitetto,
  ``A new light on the darkest corner of the landscape,''
  arXiv:1810.11365 [hep-th].
  %%CITATION = ARXIV:1810.11365;%%
  %1 citations counted in INSPIRE as of 06 Nov 2018


\bibitem{wasteland} 
  D.~Marsh, L.~McAllister and T.~Wrase,
  ``The Wasteland of Random Supergravities,''
  JHEP {\bf 1203}, 102 (2012)
  doi:10.1007/JHEP03(2012)102
  [arXiv:1112.3034 [hep-th]].
  %%CITATION = doi:10.1007/JHEP03(2012)102;%%
  %80 citations counted in INSPIRE as of 21 Oct 2018

\bibitem{Liddle-Barrow} 
  A.~R.~Liddle, P.~Parsons and J.~D.~Barrow,
  ``Formalizing the slow roll approximation in inflation,''
  Phys.\ Rev.\ D {\bf 50}, 7222 (1994)
  doi:10.1103/PhysRevD.50.7222
  [astro-ph/9408015].
  %%CITATION = doi:10.1103/PhysRevD.50.7222;%%
  %466 citations counted in INSPIRE as of 21 Oct 2018

\bibitem{future} Work in Progress.

\bibitem{Faulkner} 
  T.~Faulkner, R.~G.~Leigh, O.~Parrikar and H.~Wang,
  ``Modular Hamiltonians for Deformed Half-Spaces and the Averaged Null Energy Condition,''
  JHEP {\bf 1609}, 038 (2016)
  doi:10.1007/JHEP09(2016)038
  [arXiv:1605.08072 [hep-th]].
  %%CITATION = doi:10.1007/JHEP09(2016)038;%%
  %82 citations counted in INSPIRE as of 31 Dec 2018

\bibitem{PhD} 
  G.~Dibitetto,
  ``Gauged Supergravities and the Physics of Extra Dimensions,''
  arXiv:1210.2301 [hep-th].
  %%CITATION = ARXIV:1210.2301;%%
  %6 citations counted in INSPIRE as of 21 Oct 2018

\bibitem{Denef}
  F.~Denef, A.~Hebecker and T.~Wrase,
  ``de Sitter swampland conjecture and the Higgs potential,''
  Phys.\ Rev.\ D {\bf 98}, no. 8, 086004 (2018)
  doi:10.1103/PhysRevD.98.086004
  [arXiv:1807.06581 [hep-th]].

\bibitem{Conlon}
J.~P.~Conlon,
  ``The de Sitter swampland conjecture and supersymmetric AdS vacua,''
  arXiv:1808.05040 [hep-th].
%\cite{Kinney:2018nny}

\bibitem{Yamazaki} 
  H.~Murayama, M.~Yamazaki and T.~T.~Yanagida,
  ``Do We Live in the Swampland?,''
  arXiv:1809.00478 [hep-th].
  %%CITATION = ARXIV:1809.00478;%%
  %14 citations counted in INSPIRE as of 21 Oct 2018

\bibitem{DenefLectures} 
  F.~Denef,
  ``Les Houches Lectures on Constructing String Vacua,''
  Les Houches {\bf 87}, 483 (2008)
  [arXiv:0803.1194 [hep-th]].
  %%CITATION = ARXIV:0803.1194;%%
  %331 citations counted in INSPIRE as of 06 Nov 2018

%\cite{Liu:2002yd}
\bibitem{Seiberg} 
  H.~Liu, G.~W.~Moore and N.~Seiberg,
  ``The Challenging cosmic singularity,''
  gr-qc/0301001.
  %%CITATION = GR-QC/0301001;%%
  %22 citations counted in INSPIRE as of 21 Oct 2018

\bibitem{Horo} 
  G.~T.~Horowitz and J.~Polchinski,
  ``Instability of space - like and null orbifold singularities,''
  Phys.\ Rev.\ D {\bf 66}, 103512 (2002)
  doi:10.1103/PhysRevD.66.103512
  [hep-th/0206228].
  %%CITATION = doi:10.1103/PhysRevD.66.103512;%%
  %193 citations counted in INSPIRE as of 21 Oct 2018

%\cite{Fischler:2001yj}
\bibitem{Willy} 
  W.~Fischler, A.~Kashani-Poor, R.~McNees and S.~Paban,
  ``The Acceleration of the universe, a challenge for string theory,''
  JHEP {\bf 0107}, 003 (2001)
  doi:10.1088/1126-6708/2001/07/003
  [hep-th/0104181].
  %%CITATION = doi:10.1088/1126-6708/2001/07/003;%%
  %274 citations counted in INSPIRE as of 13 Jul 2018 

\bibitem{Susskind} 
  S.~Hellerman, N.~Kaloper and L.~Susskind,
  ``String theory and quintessence,''
  JHEP {\bf 0106}, 003 (2001)
  doi:10.1088/1126-6708/2001/06/003
  [hep-th/0104180].
  %%CITATION = doi:10.1088/1126-6708/2001/06/003;%%
  %275 citations counted in INSPIRE as of 13 Jul 2018  

  
   
 

  

  
\bibitem{Plefka} 
  G.~Jorjadze, J.~Plefka and J.~Pollok,
  ``Bosonic String Quantization in Static Gauge,''
  J.\ Phys.\ A {\bf 45}, 485401 (2012)
  doi:10.1088/1751-8113/45/48/485401
  [arXiv:1207.4368 [hep-th]].
  %%CITATION = doi:10.1088/1751-8113/45/48/485401;%%
  %10 citations counted in INSPIRE as of 17 Oct 2018

\bibitem{CK-HS} 
  C.~Krishnan and S.~Roy,
  ``Higher Spin Resolution of a Toy Big Bang,''
  Phys.\ Rev.\ D {\bf 88}, 044049 (2013)
  doi:10.1103/PhysRevD.88.044049
  [arXiv:1305.1277 [hep-th]].
  %%CITATION = doi:10.1103/PhysRevD.88.044049;%%
  %20 citations counted in INSPIRE as of 21 Oct 2018

C.~Krishnan, A.~Raju and S.~Roy,
  ``A Grassmann path from $AdS_3$ to flat space,''
  JHEP {\bf 1403}, 036 (2014)
  doi:10.1007/JHEP03(2014)036
  [arXiv:1312.2941 [hep-th]].
  %%CITATION = doi:10.1007/JHEP03(2014)036;%%
  %49 citations counted in INSPIRE as of 22 Oct 2018

  C.~Krishnan and S.~Roy,
  ``Desingularization of the Milne Universe,''
  Phys.\ Lett.\ B {\bf 734}, 92 (2014)
  doi:10.1016/j.physletb.2014.05.039
  [arXiv:1311.7315 [hep-th]].
  %%CITATION = doi:10.1016/j.physletb.2014.05.039;%%
  %18 citations counted in INSPIRE as of 21 Oct 2018

  B.~Craps, C.~Krishnan and A.~Saurabh,
  ``Low Tension Strings on a Cosmological Singularity,''
  JHEP {\bf 1408}, 065 (2014)
  doi:10.1007/JHEP08(2014)065
  [arXiv:1405.3935 [hep-th]].
  %%CITATION = doi:10.1007/JHEP08(2014)065;%%
  %8 citations counted in INSPIRE as of 21 Oct 201

  K.~S.~Kiran, C.~Krishnan, A.~Saurabh and J.~Simón,
  ``Strings vs. Spins on the Null Orbifold,''
  JHEP {\bf 1412}, 002 (2014)
  doi:10.1007/JHEP12(2014)002
  [arXiv:1408.3296 [hep-th]].
  %%CITATION = doi:10.1007/JHEP12(2014)002;%%
  %10 citations counted in INSPIRE as of 21 Oct 2018

  C.~Krishnan and A.~Raju,
  ``Gauging Away a Big Bang,''
  J.\ Phys.\ Conf.\ Ser.\  {\bf 883}, no. 1, 012015 (2017)
  doi:10.1088/1742-6596/883/1/012015
  [arXiv:1504.04331 [hep-th]].
  %%CITATION = doi:10.1088/1742-6596/883/1/012015;%%
  %2 citations counted in INSPIRE as of 21 Oct 2018

\bibitem{Strominger} 
  A.~Strominger,
  ``The dS / CFT correspondence,''
  JHEP {\bf 0110}, 034 (2001)
  doi:10.1088/1126-6708/2001/10/034
  [hep-th/0106113].
  %%CITATION = doi:10.1088/1126-6708/2001/10/034;%%
  %824 citations counted in INSPIRE as of 21 Oct 2018

\bibitem{Banks} 
  T.~Banks and W.~Fischler,
  ``An Holographic cosmology,''
  hep-th/0111142.
  %%CITATION = HEP-TH/0111142;%%
  %113 citations counted in INSPIRE as of 21 Oct 201

\bibitem{Return-of-the-Vafa} 
  H.~Ooguri, E.~Palti, G.~Shiu and C.~Vafa,
  ``Distance and de Sitter Conjectures on the Swampland,''
  arXiv:1810.05506 [hep-th].
 
\end{thebibliography}
\end{document}